\documentclass[
reprint,
nofootinbib,
 amsmath,amssymb,
 aps,
 floatfix,
]{revtex4-2}

\usepackage[english]{babel} 

\usepackage{graphicx}
\usepackage{dcolumn}
\usepackage{bm}
\usepackage{orcidlink} 

\usepackage{array}
\usepackage[caption=false,font=normalsize,labelfont=sf,textfont=sf]{subfig}
\usepackage{textcomp}
\usepackage{url}
\usepackage{verbatim}
\usepackage{float}
\usepackage{upgreek}
\usepackage{mathrsfs} 

\usepackage{placeins} 

\usepackage{makecell} 

\begin{document}

\selectlanguage{english}

\setcounter{topnumber}{5}     
\setcounter{totalnumber}{10}  
\renewcommand{\textfraction}{0.1} 
\renewcommand{\floatpagefraction}{0.9} 

\preprint{APS/123-QED}

\title{Simulation and measurement of Black Body Radiation background \\ in a Transition Edge Sensor}

\author{José Alejandro Rubiera Gimeno\orcidlink{0000-0002-4012-9509}}
 \altaffiliation[Now at ]{Helmut-Schmidt-Universität (HSU), Holstenhofweg 85, 22043 Hamburg, Germany}
 \email{jose.rubiera.gimeno@desy.de}
\author{Friederike Januschek}
\author{Axel Lindner\orcidlink{0000-0001-6006-0820}}
\author{Christina Schwemmbauer\orcidlink{0000-0002-2029-1162}}
\affiliation{
 Deutsches Elektronen-Synchrotron DESY, Notkestr. 85, 22607 Hamburg, Germany
}%

\author{Katharina-Sophie Isleif\orcidlink{0000-0001-7032-9440}}%
\affiliation{%
 Helmut-Schmidt-Universität (HSU), Holstenhofweg 85, 22043 Hamburg, Germany
}

\author{Manuel Meyer\orcidlink{0000-0002-0738-7581}}
\author{Elmeri Rivasto\orcidlink{0000-0002-1255-0726}}
\affiliation{
 CP3-Origins, University of Southern Denmark, Campusvej 55, 5230 Odense M, Denmark
}%

\author{Gulden Othman}%
\affiliation{
 Institut für Experimentalphysik, Universität Hamburg UHH, Notkestr. 85, 22607 Hamburg, Germany
}%
\altaffiliation[Now at ]{Helmut-Schmidt-Universität (HSU), Holstenhofweg 85, 22043 Hamburg, Germany}

\author{Rikhav Shah\orcidlink{0000-0002-3329-9075}}
\affiliation{
 Institute for Physics, Johannes-Gutenberg-Universität JGU, Staudingerweg 7, 55128 Mainz, Germany
}%
\altaffiliation[Now at ]{Institut für Quantenphysik, Universität Hamburg UHH, Notkestr. 85, 22607 Hamburg, Germany}



\begin{abstract}

The Any Light Particle Search~II (ALPS~II) experiment at DESY, Hamburg, is a Light-Shining-through-a-Wall (LSW) experiment aiming to probe the existence of axions and axion-like particles (ALPs), which are candidates for dark matter. Data collection in ALPS~II is underway utilizing a heterodyne-based detection scheme. A complementary run for confirmation or as an alternative method is planned using single photon detection, requiring a sensor capable of measuring low-energy photons (1064\,nm, 1.165\,eV) with high efficiency (higher than 50\,\%) and a low background rate (below $7.7\cdot10^{-6}\,\mathrm{cps}$). 
To meet these requirements, we are investigating a tungsten Transition Edge Sensor (TES) provided by NIST, which operates in its superconducting transition region at millikelvin temperatures. This sensor exploits the drastic change in resistance caused by the absorption of a single photon.
We find that the background observed in the setup with a fiber-coupled TES is consistent with Black Body Radiation (BBR) as the primary background contributor.
A framework was developed to simulate BBR propagation to the TES under realistic conditions.
The framework not only allows the exploration of background reduction strategies, such as improving the TES energy resolution, but also reproduces, within uncertainties, the spectral distribution of the observed background.
These simulations have been validated with experimental data, in agreement with the modeled background distribution, and show that the improved energy resolution reduces the background rate in the 1064\,nm signal region by one order of magnitude, to approximately $10^{-4}\,\mathrm{cps}$.
However, this rate must be reduced further to meet the ALPS II requirements.

\end{abstract}

\maketitle



\section{Introduction}
\label{sec:intro}

Axions and axion-like particles (ALPs) are hypothetical pseudo-scalar bosons proposed in theories beyond the Standard Model \cite{axionboson}. Light-shining-through-a-wall (LSW) experiments~\cite{anselm_arion-photon_1985, Bibber_lsw} constitute a model-independent method to search for these particles by probing their interaction with photons, characterized by the coupling strength $g_{a\gamma \gamma}$. In LSW experiments, photons are converted into ALPs within a magnetic field (Primakoff effect) and back into photons of the same energy as the primary within another magnetic field (Sikivie effect)~\cite{anselm_arion-photon_1985, Bibber_lsw}.

The Any Light Particle Search~II~(ALPS~II) experiment at Deutsches Elektronen-Synchrotron (DESY)~\cite{alpsiitechnicaldesign} is a second-generation LSW experiment. It is designed to improve sensitivity to the coupling $g_{a\gamma \gamma}$ by a factor of $10^{3}$ compared to earlier experiments~\cite{alps2sensitivity, OSQAR:2015qdv} through the use of high-finesse optical cavities and a longer magnet string, targeting $g_{a\gamma\,\gamma}\,=\,2\,\cdot 10^{-11}\,\mathrm{GeV^{-1}}$. This target is motivated by astrophysical anomalies, such as stellar evolution~\cite{stellarevolution}.
Within the ALPS~II experiment, for a 1064~nm photon rate equivalent to 30\,W of laser power and $g_{a\gamma\,\gamma} = 2\,\cdot\,10^{-11}\,\mathrm{GeV^{-1}}$, a signal from reconverted photons of $10^{-24}\,\mathrm{W}$, or about $10^{-5}\,\mathrm{cps}$—roughly one photon per day, would be expected~\cite{alps2counts}.
To detect these photons, two methods have been investigated: heterodyne (HET)~\cite{alps2het} and single photon detection~\cite{alpsiitechnicaldesign, epsrikhav}. ALPS~II is currently using HET for data taking, with single photon counting detection as a confirmation or alternative sensing scheme having very different systematics.

For a single photon detection scheme, the system must detect photons with $1.165\,\mathrm{eV}$ energy at $10^{-5}\,\mathrm{cps}$. Additionally, to achieve $5\sigma$ significance in ALP detection within 20 days, the detector needs over 50~\% efficiency and a background rate below $7.7\cdot10^{-6}\,\mathrm{cps}$ \cite{alpsiitechnicaldesign, epsrikhav}. A Transition Edge Sensor (TES) is being studied as a strong candidate to meet these requirements.
This paper focuses on the minimization of background rates, particularly from Black Body Radiation (BBR).
We present the development of a simulation framework for modeling BBR propagation to the TES and evaluate the contribution of BBR to the overall TES background. With the simulation framework, we are also able to model the impact of the strategies for background reduction.
We discuss the results of the ongoing efforts to enhance the TES energy resolution through the optimization of the data analysis, which impacts the background rate measured with the TES.

This paper is organized into six sections. Section~\ref{sec:tes_setup} provides details of the TES measurement setup. Section~\ref{sec:BBR_sim} describes the theoretical and experimental aspects of the BBR simulation framework, which models the propagation of Black Body Radiation to the TES. Section~\ref{sec:example} demonstrates an example application of the framework, adapted to the specific experimental conditions outlined in Section~\ref{sec:extrinsiccompartison}. In Section~\ref{sec:extrinsiccompartison}, the analysis of measured background data is explained. Furthermore, a comparison between simulation predictions and experimental results is made. Finally, the key conclusions of this paper and the next steps for future work are presented in Section~\ref{sec:conclusion}.

\section{TES setup}
\label{sec:tes_setup}

A Transition Edge Sensor (TES) is a cryogenic microcalorimeter that operates in the transition region between superconducting and normal conducting states~\cite{Irwin2005}. It responds to energy depositions with an abrupt change in resistance, generating a measurable signal. In our setup, the TES consists of a tungsten microchip provided by the National Institute of Standards and Technology (NIST), USA. The TES has dimensions of $25~\mathrm{\upmu m} \times 25~\mathrm{\upmu m} \times 20~\mathrm{nm}$, a critical temperature around $140~\mathrm{mK}$, and is optimized for 1064~nm photon detection.
The signal produced in the TES due to energy deposition is measured as a voltage signal using a readout system based on a two-stage Superconducting Quantum Interference Device (SQUID) provided by Physikalisch-Technische Bundesanstalt (PTB), Berlin, Germany, and SQUID electronics from Magnicon.
The TES and the SQUID cold components are housed within a Bluefors dilution refrigerator operating at a base temperature of $25~\mathrm{mK}$. For more details on the setup, see~\cite{jltprikhav}.

Photons are guided to the TES via an optical fiber from outside the cryostat. To characterize the TES response at the ALPS~II operating wavelength, light from a 1064~nm laser is introduced into the system, highly attenuated to avoid saturating the sensor. This is adjusted using several attenuators to minimize pile-up effects in the data~\cite{rubiera_gimeno_tes_2023}.

In the \emph{extrinsics} measurement, the background is assessed by connecting the TES to the optical fiber while leaving the other end of the fiber covered at room temperature~\cite{rubiera_gimeno_tes_2023}. Under these conditions, using the same acquisition settings as in the 1064~nm laser measurements, a trigger rate on the order of $10^{-2}~\mathrm{cps}$ is observed. This rate is expected to be primarily due to BBR from the laboratory environment coupling into the fiber~\cite{miller2007superconducting}. A framework to compute the expected contribution of BBR will be described in the following.

\section{A framework to simulate BBR propagating to the TES}
\label{sec:BBR_sim}

We have developed a Python framework to compute the BBR photons propagating through the optical fiber to the TES. The model for the simulation is built by considering the theory of BBR production, optical components in the TES setup acting like a filter, and the TES response.
The individual parts of the simulation will be explained in this section.

\subsection{Production of Black Body Radiation}
\label{BBR_production}

A perfect black body emits photons with energy~$E$ according to Planck's distribution~\cite{greiner_thermodynamics_1995}. The spectral intensity, which depends on the photon energy and the temperature~$T$, is given by
\begin{equation}
\label{eq:BBR}
    B(E, T) = \frac{2}{h^3 c^2} \frac{E^2}{\exp\left(\frac{E}{kT}\right) - 1},
\end{equation}
where $B(E, T)$ represents the rate of photons emitted per unit area, energy and solid angle, measured in $\mathrm{photon/(eV\cdot s\cdot m^2 \cdot sr)}$. Here, $h$ is Planck's constant, $c$ is the speed of light, and $k$ is the Boltzmann constant. The angular distribution of the emitted photons follows Lambert's cosine law, meaning that the emission rate in an infinitesimal solid angle is proportional to $B(E, T) \cos{\theta}$, where $\theta$ is the angle between the surface normal and the direction of emission.

For an optical fiber, the flux of photons coupling into one end is equal to the flux that the fiber’s surface would emit. This flux is determined by the surface area~$A$ of the fiber and the coupling efficiency~$\eta_e(E, \theta)$, which depends on the photon's energy~$E$ and the incidence angle~$\theta$. The photon flux coupled into the fiber can be computed as
\begin{equation}
\label{eq:perfect_BBR_rate}
    \frac{d}{dE} N (E, T) = \iint_{A, \Omega} B(E, T) \eta_e (E, \theta) \cos{\theta} \, d\Omega \, dA,
\end{equation}
where the integral is taken over the surface area~$A$ and the solid angle~$\Omega$ and $N (E, T)$ is the photon rate.

\subsection{Energy dependent transfer function of optical components}
\label{sec:opt_components}

\textit{TES Absorbance:}
The quantum efficiency of the TES under study is optimized for a wavelength of 1064~nm through the inclusion of a cavity-like coating structure around the sensor~\cite{lita_opt_stack}. This optimization allows the TES to achieve an efficiency of up to 99.68~\% at 1064~nm, as shown in~\cite{janthesis}.

For computational simplicity, the following expression is used as an approximation of the efficiency curve:
\begin{equation}
\label{eq:TES_absorbance}
    \begin{gathered}
    1 - \eta_{\mathrm{TES}}(E) = f_1 (E) \cdot f_2 (E)  \\
    f_1 (E) = \frac{1 - b_1}{2}\left( 1 + \text{erf} \left \{
                a_1 \left(\lambda(E) - \lambda_1\right) \right 
            \} \right)  + b_1    \\
    f_2 (E) = \frac{1 - b_2}{2}\left( 1 - \text{erf} \left \{
                a_2 \left(\lambda(E) - \lambda_2\right) \right 
            \} \right)  + b_2.
    \end{gathered}
\end{equation}

\begin{figure}[!h]
    \centering
    \includegraphics[width=0.35\textwidth]{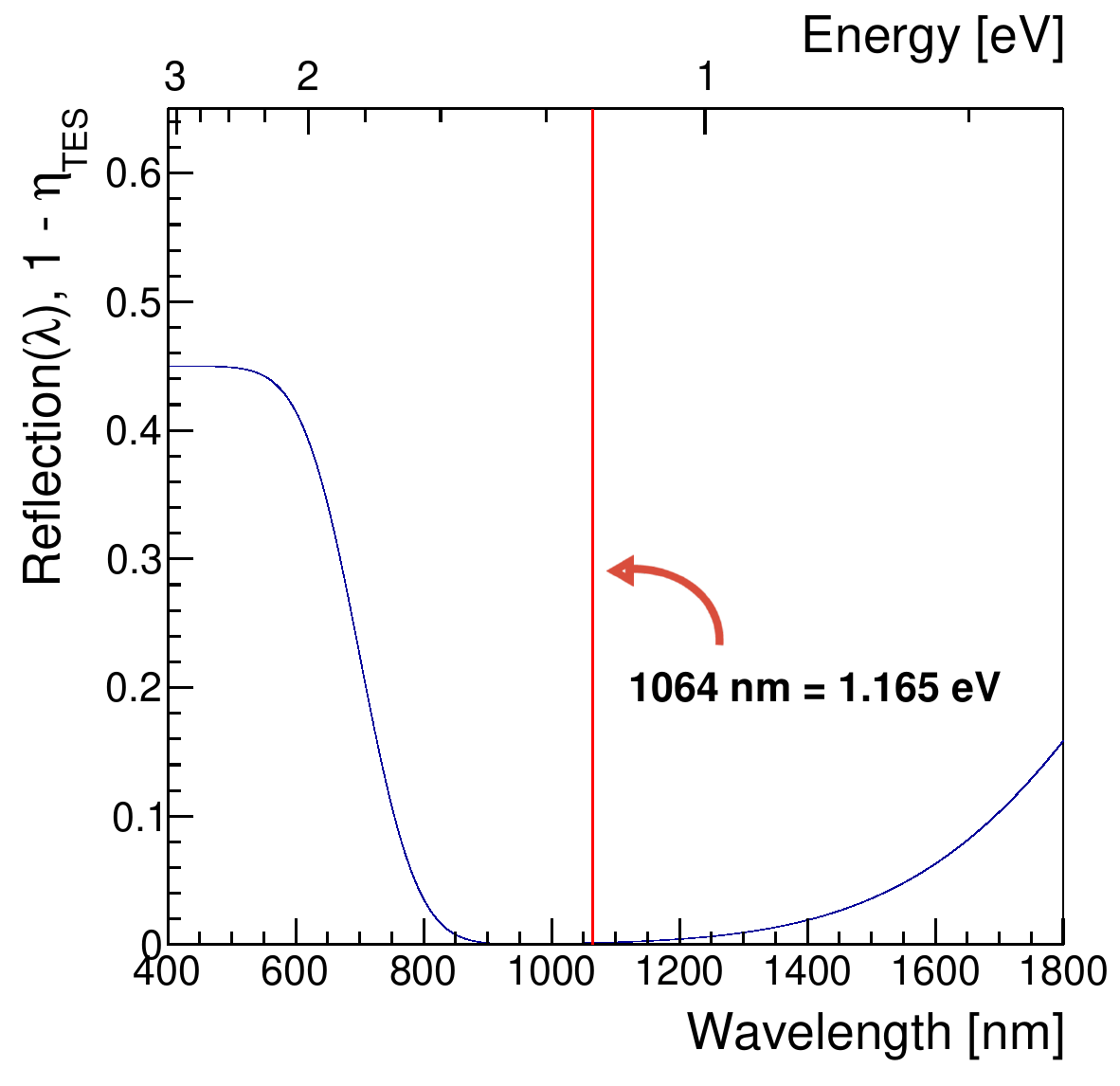}    
    \vspace{-10pt}     
    \caption{
        Model of the TES reflectance as a function of the wavelength as implemented in the BBR simulation.
    }
    \label{fig:simulation_TES_reflexion}
\end{figure}

Equation~\ref{eq:TES_absorbance} is inspired by the reflection curve depending on the wavelength in~\cite{janthesis}, and $1~-~\eta_{\mathrm{TES}}(E)$ represents the reflection fraction, while $\eta_{\mathrm{TES}}(E)$ denotes the absorbance of the TES. To approximate the TES reflection curve, the error functions in $f_1$ and $f_2$ are combined, acting as a band-pass filter, as shown in Fig.~\ref{fig:simulation_TES_reflexion}. 

This model assumes that the reflection fraction will exhibit a minimum at the wavelength where the TES is optimized.
In Eq.~\ref{eq:TES_absorbance}, $1 - \eta_{\mathrm{TES}}(E)$ converges to $1 - b_1$ at high energies (short wavelengths) and $1 - b_2$ at low energies (long wavelengths), as shown in Fig.~\ref{fig:simulation_TES_reflexion}. The parameters $a_1$ and $a_2$ adjust the steepness of the error functions, while $\lambda_1$ and $\lambda_2$ determine the central positions of these transitions. 

\begin{table}[!h]
    \centering
    \caption{
    TES reflectance parameters.
    }
    \begin{tabular}{|c|c|c|c|c|c|}    
    \hline
        $\lambda_1$~[nm] & $b_1$ & $a_1$~[1/nm] & $\lambda_2$~[nm] & $b_2$ & $a_2$~[1/nm] \\
    \hline
        700 & 0.55 & 0.01 & 2100 & 0.2 & 0.002 \\
    \hline
    \end{tabular}
    \label{tab:reflection_parameters}
\end{table}

The parameters shown in Table~\ref{tab:reflection_parameters} are selected to ensure that the modeled reflection curve (Fig.~\ref{fig:simulation_TES_reflexion}) closely resembles the reflection data in~\cite{janthesis}.

Even though the approximation shows some deviations from the measured data in~\cite{janthesis}, it does not
introduce a significant error in the simulation (only higher than 10~\% at wavelengths lower than 650~nm). This is due to the fact that at wavelengths below 650~nm, the rate of BBR photons is lower by several orders of magnitude compared to the rate at 1064~nm due to the exponential behavior described by Eq.~\ref{eq:BBR}. For wavelengths above 1800~nm, there is insufficient data to accurately model TES absorbance. Nonetheless, BBR photons with these longer wavelengths are significantly suppressed by optical fiber curling, as will be discussed in the next sections. Therefore, deviations in the approximation in Eq.~\ref{eq:TES_absorbance} do not have a relevant impact on the simulation results.

\begin{figure}[!h]
    \centering
    \captionsetup[subfloat]{captionskip=-12pt}
    \subfloat[]{
        \includegraphics[width=0.8\columnwidth]{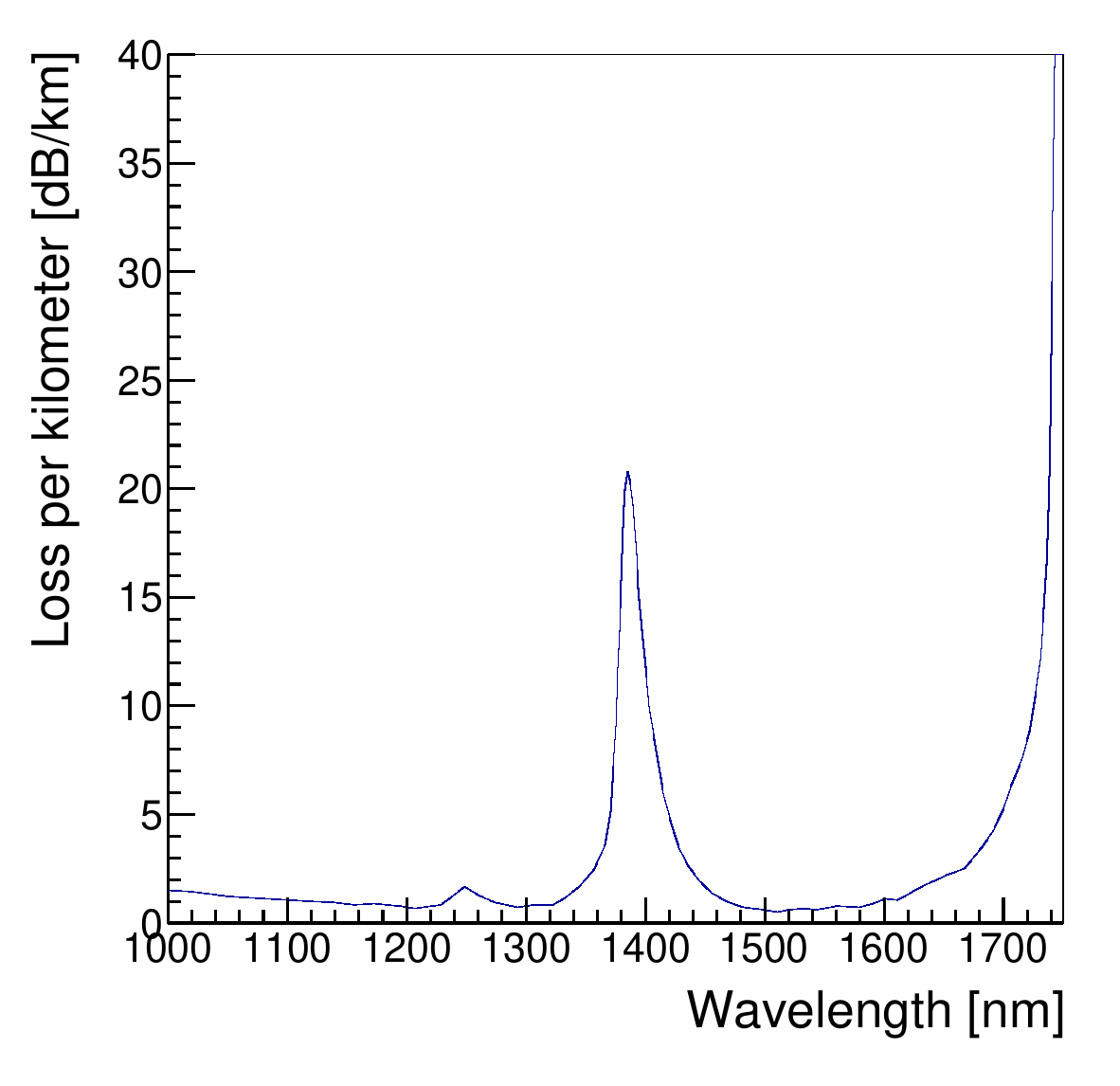}
        \label{fig:simulation_fiber_loss}
    }    
    
    \subfloat[]{
        \includegraphics[width=0.8\columnwidth]{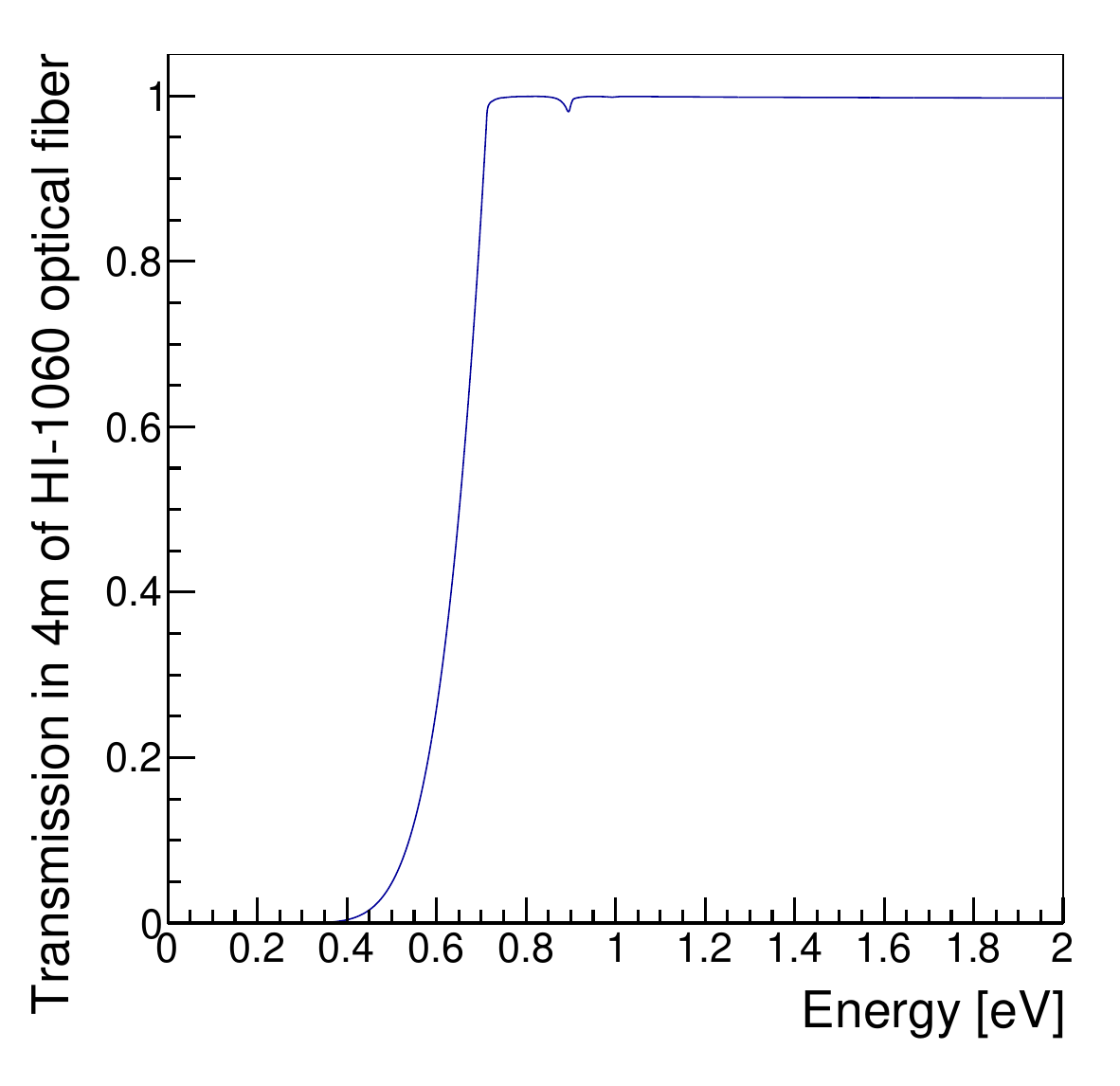}
        \label{fig:simulation_fiber_transmission}
    }
    \caption{
        (a) Loss per kilometer of a HI-1060 optical fiber from Corning, the data was taken from~\cite{fiber_loss}, (b) Transmission of 4~m of optical fiber as a function of the photon energy computed from the loss per kilometer.
    }
    \label{fig:simulation_fiber}
\end{figure}

\textit{Fiber transmission:}
The fiber transmission is calculated based on the loss per kilometer of optical fiber, which varies with the wavelength of the guided photons. This calculation does not take into account coupling or bending losses. To minimize the loss of 1064~nm photons, a HI-1060 single-mode optical fiber from Corning~\cite{fiber_loss} is used in the TES setup. The loss per kilometer for this fiber is shown in Fig.~\ref{fig:simulation_fiber_loss}. For wavelengths outside the range displayed on the x-axis, the losses are extrapolated using a linear function. 

The transmission $\eta_{\mathrm{fiber}}(E, l)$, as a function of the energy of the propagating photon $E$, is then computed for a given fiber length $l$ in meters. Fig.~\ref{fig:simulation_fiber_transmission} illustrates the transmission for a 4~m length of the HI-1060 fiber.

\begin{figure}[!h]
    \centering
    \includegraphics[width=0.8\columnwidth]{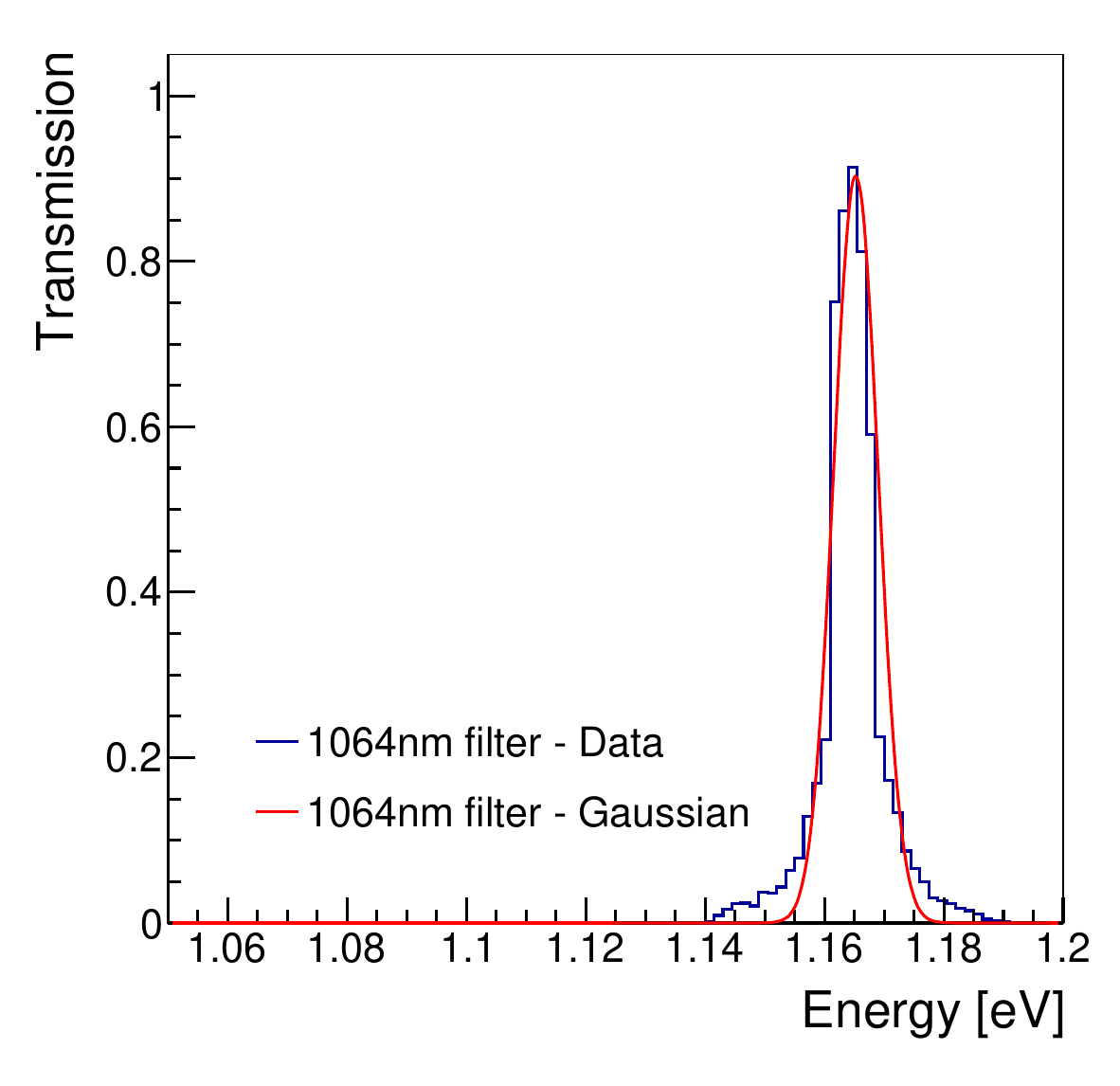}    
    \vspace{-10pt}     
    \caption{
        Transmission data as a function of energy from the datasheet of a FLH1064-8 filter from Thorlabs. The peak in the data as a function of wavelength is centered at 1064~nm with a FWHM of 8~nm. The data in the datasheet is in the form of transmission vs. wavelength. The Gaussian function describing the filter is computed with a maximum equal to the maximum transmission in the datasheet, $\mu = 1064~\mathrm{nm}$ and $\sigma = 8~\mathrm{nm}/2.355$. The Gaussian function is plotted as a function of energy instead of wavelength. 
    }
    \label{fig:simulation_optical_filter}
\end{figure}

\textit{Optical filter:}
The use of a reflective filter at cryogenic temperatures is proposed in \cite{miller2007superconducting} as a strategy to reduce the background generated by BBR. This component has been incorporated into the simulation to evaluate its potential application in the TES system for ALPS~II, specifically for filtering out non-1064~nm photons within the cold environment of the cryostat. The filter's transmission, $\eta_{\mathrm{filter}}(E)$, can be derived from the filter's datasheet or approximated using a Gaussian distribution. Fig.~\ref{fig:simulation_optical_filter} presents the simulated transmission distribution for the FLH1064-8 filter from Thorlabs, using both the datasheet information and the Gaussian approximation. This filter is characterized by a central wavelength of 1064~nm and a Full Width at Half Maximum (FWHM) of 8~nm.

\textit{Fiber curling:}
The bending of optical fibers introduces wavelength-dependent losses, which increase with the wavelength and depend on the fiber curvature radius~\cite{Simplified_Bending_loss_theory}. Consequently, fiber curling can be employed in a cryogenic environment to selectively suppress the lower-energy portion of the BBR spectrum. This technique was explored, for example, in \cite{Smirnov_2015}, where fiber curling at low temperatures was used to reduce BBR rates, measured using a superconducting nanowire single-photon detector (SNSPD).

\begin{figure}[!h]
    \centering
    \includegraphics[width=0.8\columnwidth]{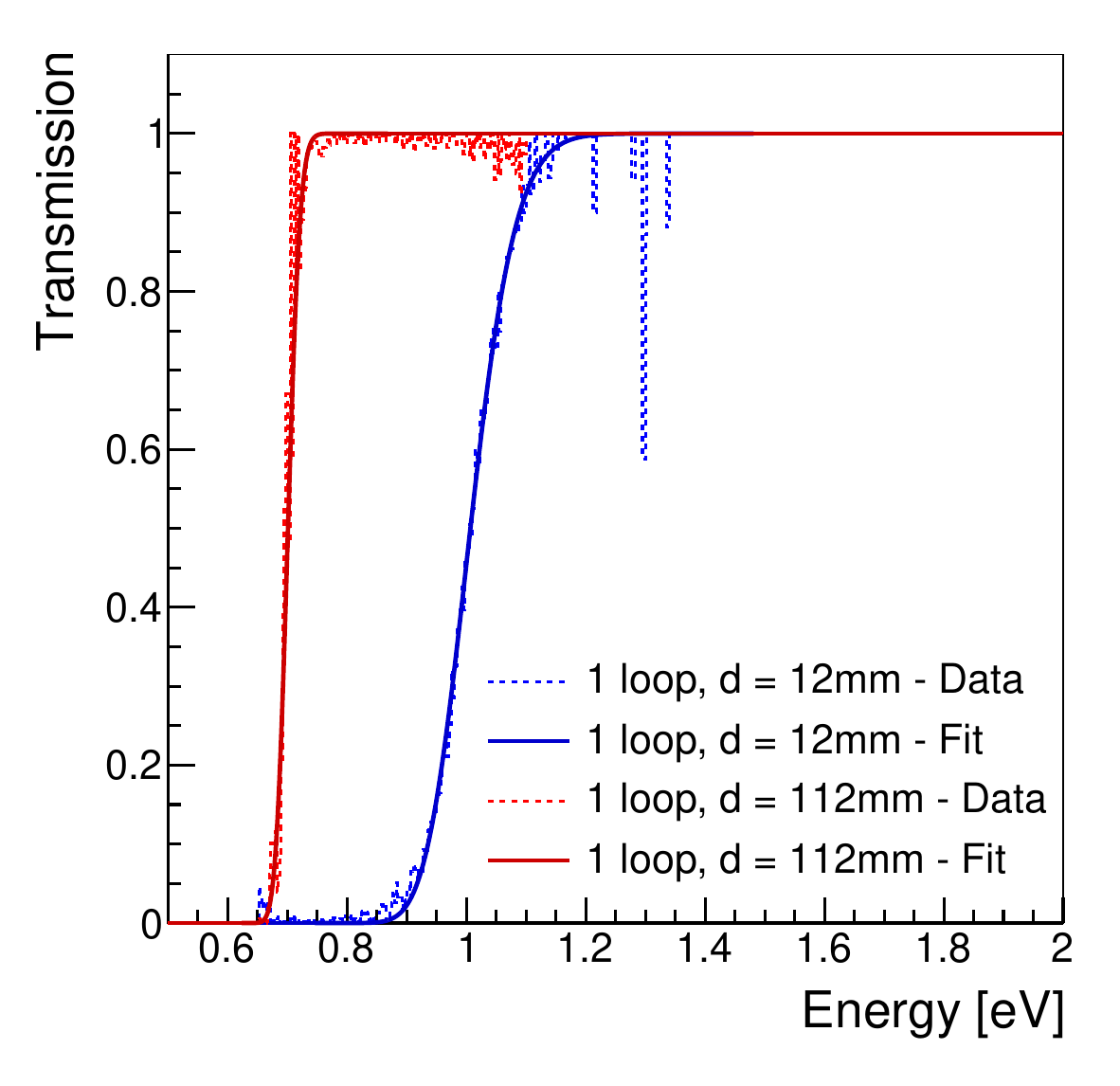}    
    \vspace{-10pt}     
    \caption{
        Example of the transmission as a function of energy for one loop of curled fiber with diameters $d = 1.2~\mathrm{cm}$ and $d = 11.2~\mathrm{cm}$ measured with a spectrometer. The data is fitted to an error function for use in the BBR simulation.
    }
    \label{fig:simulation_fiber_curling}
\end{figure}

In this study, we utilized empirical data, shown in Fig.~\ref{fig:simulation_fiber_curling}, which characterizes fiber loss as a function of wavelength and bending radius for a single loop. White light from a Thorlabs SLS201L lamp was used as a source of continuous spectrum connected to an optical fiber. The other end of the optical fiber was connected to an Ocean Insight NQ512-2.5 spectrometer. The spectrum of transmission shown in Fig.~\ref{fig:simulation_fiber_curling} was computed as the ratio between the spectrum measured when the fiber was curled, under controlled curling diameter, and the spectrum measured when the fiber was not curled. 
This empirical data was fitted with a phenomenological model described by
\begin{equation}
\label{eq:TES_curling}
    \eta_{\mathrm{curl}}(E, \alpha) = \left[ \frac{1}{2}\left( 1 + \text{erf} \left \{
                a (\lambda(E) - \lambda_0) \right \} \right)
                \right]^\alpha,
\end{equation}
 and integrated into the simulation of BBR. Eq.~\ref{eq:TES_curling} was chosen to reproduce the characteristic sharp decrease in transmission due to bending losses, while providing a smooth and flexible approximation of the transmission as a function of the wavelength, as illustrated in Fig.~\ref{fig:simulation_fiber_curling}. Here, $E$ denotes the photon's energy. The exponent $\alpha$ was introduced to account for multiple loops in the fiber. Since the total transmission through multiple independent loops can be modeled as the product of the individual transmissions, $\alpha$ represents the number of loops, with fractional values indicating partial loops.

\begin{table}[!h]
    \centering
    \vspace{-5pt}
    \caption{
    One-loop fiber curling parameters.
    }
    \label{tab:curling_parameters}
    \begin{tabular}{|c|c|c|}
    \hline
        Curling diameter [mm] & $a$~[1/nm] & $\lambda_0$~[nm]\\
    \hline
        12.0 & $9.71 \cdot 10^{-3}$ & 1231 \\
    \hline
        25.2 & $7.96 \cdot 10^{-3}$ & 1378 \\
    \hline
        42.0 & $1.22 \cdot 10^{-2}$ & 1598 \\
    \hline
        112.0 & $1.75 \cdot 10^{-2}$ & 1768 \\
    \hline
    \end{tabular}
\end{table}

The parameters $a$ and $\lambda_0$ shown in Table~\ref{tab:curling_parameters} are determined by the width and center of the transmission transition for a single loop of curled fiber, respectively.

\FloatBarrier

\subsection{Combination of BBR and optical components}
\label{sec:combination}

Considering the optical components and the production of BBR, a model can be built to compute the rate of photons arriving at the TES. Taking into account every path $j$ that the BBR photons with energy $E$ would need to follow to reach the detector, we use the following expression:
\begin{equation}
\label{eq:BBR_components}
    \begin{gathered}
        \frac{d}{dE} N (E) = \sum_{j} \iint_{A_j, \Omega_j} B_j(E, T_j) \cos{\theta} \, d\Omega \, dA \\
        B_j = B(E, T_j)  \cdot \prod_{i_j} \eta_{i_j} \cdot H(\theta_{\mathrm{max}_j} - \theta).
    \end{gathered}
\end{equation}

The starting point of each path $j$ corresponds to where BBR can couple into the optical system, such as at fiber junctions, e.g., fiber mating sleeves. Depending on the experimental setup and optical path, this could include paths $j$ with multiple starting points for the same setup. Then, for every path $j$, the product operator multiplies the transmissions $\eta_{i_j}$ of every optical component $i_j$ along it to the TES. Moreover, a Heaviside step function $H(\theta_{\mathrm{max}_j} - \theta)$ introduces the approximation that the optical fiber, used to transmit the BBR to the TES, accepts every photon entering with an angle $\theta$ lower than $\theta_{\mathrm{max}_j}$ with respect to the central axis of the core of the optical fiber, and rejects the rest. Furthermore, it assumes that the transmission of the optical components does not depend on $\theta$.

The advantage of the approach in Eq.~\ref{eq:BBR_components} is its flexibility, which allows the inclusion of different combinations of optical components depending on the experimental conditions.

\subsection{TES response - Energy and time resolution}
\label{sec:combination_convolution}

Equation~\ref{eq:BBR_components} would be the spectrum of BBR measured by the TES if it had perfect energy and time resolution. To adapt it to the real conditions, the TES response to a photon flux needs to be considered. To account for this, we fold $\frac{dN(E)}{dE}$ with the TES response.

The spectra measured by the TES from a single energy source can be approximated to follow a Gaussian distribution $\mathscr{N}(\epsilon, \mu = E, \sigma = \sigma(E))$, with $\epsilon$ being the measured energy, $\mu$ being equivalent to the true energy of the photon $E$, and $\sigma$ given by the energy resolution, which can be energy dependent. The energy resolution at 1064\,nm, $\sigma_{\mathrm{1064nm}}$, is taken as a reference, and it is assumed that the energy resolution is constant for other energies such that $\sigma(E) = \sigma_{\mathrm{1064nm}}$. In the energy range of interest 
[0\,eV, 1.5\,eV],
this is motivated by our previous measurements and results from other groups~\cite{cabrera_linear_tes, enrgres, multicolor}. Equation~\ref{eq:direct_BBR_rate} convolves the rate $N(E)$ and the TES energy resolution to describe the rate of single photons, $\hat{N}_{\mathrm{dir}}$, measured at the TES ("direct\footnote{In contrast to indirect events which refers to pile up events as discussed next} at 1064~nm"):
\begin{equation}
\label{eq:direct_BBR_rate}
    \frac{d}{dE} \hat{N}_{\mathrm{dir}} (\epsilon) = \frac
                {\int_{0}^{\infty} \frac{d}{dE} N (E) \mathscr{N}(\epsilon, E, \sigma(E)) \, dE}
                {\int_{0}^{\infty} \mathscr{N}(\epsilon, E, \sigma(E)) \, dE} .
\end{equation}

The process is equivalent to a convolution of the BBR spectrum with the response of the TES. The denominator in Eq.~\ref{eq:direct_BBR_rate} is added to keep the total BBR rate invariant.

Due to the limited time resolution of the TES, another effect that needs to be considered is the arrival of two photons with a time difference $\Delta t$, where the TES voltage has not yet returned to its baseline. This effect is defined as pile-up. If $\Delta t$ is sufficiently small, the pulses produced by the photons cannot be distinguished and could mimic a single photon, with energies piled-up, an effect that is often used for measuring photon number distribution with TESs~\cite{lita_opt_stack}. A value of $\Delta t_{\mathrm{min}} = 0.75~\mathrm{\upmu s}$ is assumed to be the minimum that our current data analysis can distinguish. The possible BBR spectrum from the pile-up contribution, $N_p$, motivated by~\cite{Pileup_BBR}, is computed using Eq.~\ref{eq:unfolded_indirect_BBR_rate}:
\begin{equation}
\label{eq:unfolded_indirect_BBR_rate}
    \frac{d}{dE} N_{p} (\epsilon) = 2\Delta t_{\mathrm{min}} \int_{0}^{\epsilon} \frac{d}{dE} N(E) \frac{d}{dE} N(\epsilon - E) \, dE.
\end{equation}

Equation~\ref{eq:unfolded_indirect_BBR_rate} is inspired from the probability of two events with energies $E$ and $\epsilon - E$ resulting in an event with energy $\epsilon$, occurring with a time difference lower than $\Delta t_{\mathrm{min}}$. The upper limit in the integral avoids $\epsilon - E$ to be negative.

This is then convolved with the energy resolution ("indirect" at 1064~nm as pile-up):
\begin{equation}
\label{eq:indirect_BBR_rate}
    \frac{d}{dE} \hat{N}_{\mathrm{ind}} (\epsilon) = \frac
                {\int_{0}^{\infty} \frac{d}{dE} N_{p} (E) \mathscr{N}(\epsilon, E, \sigma(E)) \, dE}
                {\int_{0}^{\infty} \mathscr{N}(\epsilon, E, \sigma(E)) \, dE}.
\end{equation}

The total rate produced by the BBR is given by adding the single photons and the pile-up contributions:
\begin{equation}
\label{eq:total_folded_BBR_rate}
    \frac{d}{dE}\hat{N}_{\mathrm{total}}(E) = \frac{d}{dE}\hat{N}_{\mathrm{dir}}(E) + \frac{d}{dE}\hat{N}_{\mathrm{ind}}(E).
\end{equation}

\section{An example built from the measurement conditions}
\label{sec:example}

\begin{figure}[h!]
    \centering
        \vspace{ 6pt}
        \includegraphics[width=0.9\columnwidth]{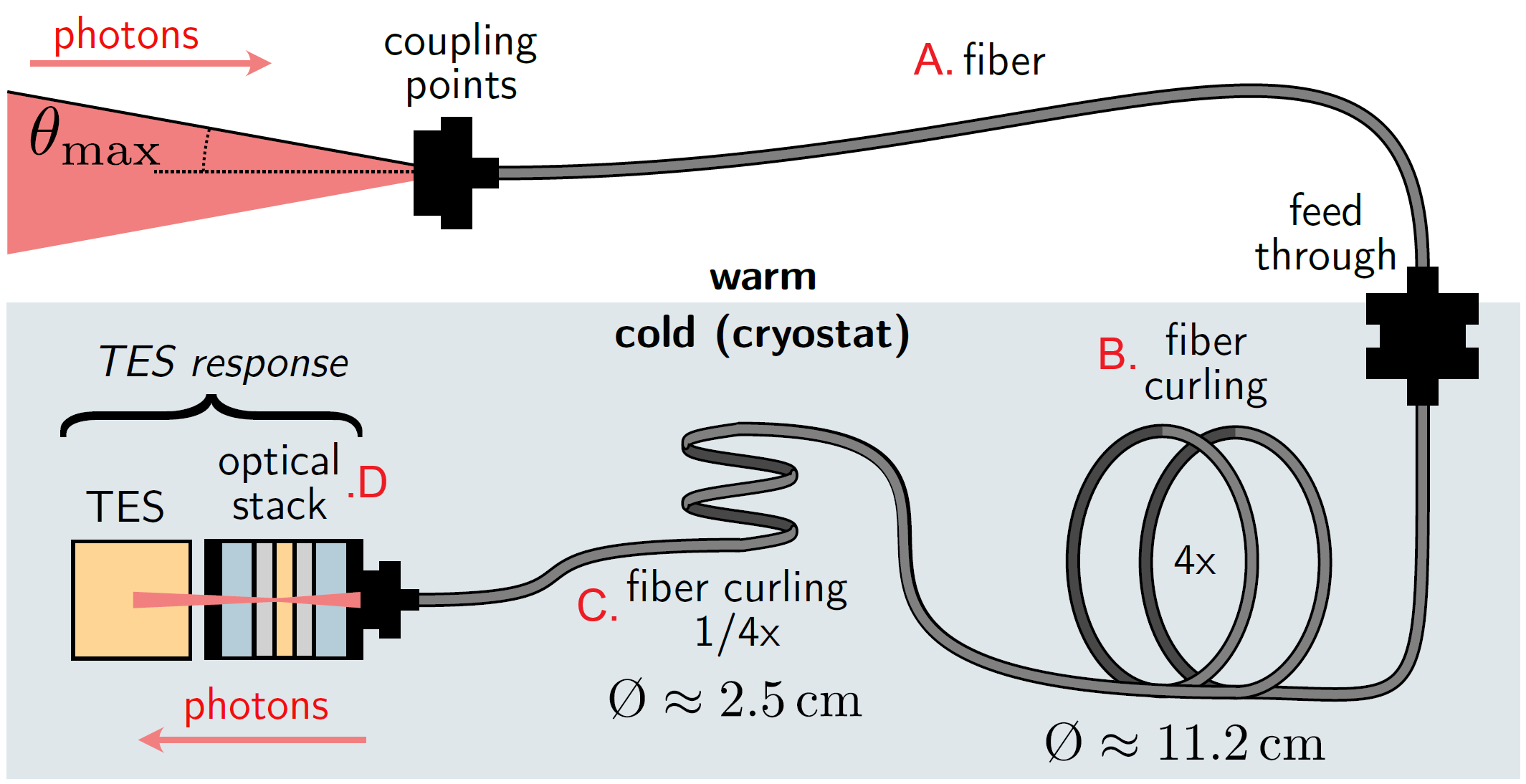}
        \vspace{-6pt}        
    \caption{
        Scheme of the simulated path in the BBR simulation. The BBR enters the optical fiber in the cone on the left of (A) if the incidence angle $\theta$ with respect to the fiber core axis is lower than $\theta_{\mathrm{max}}$ given by the numerical aperture of the fiber. (A) represents the optical fiber outside the cryostat. The BBR absorbed by the fiber is emitted back inside the fiber because it is at room temperature. The result is equivalent to directly producing BBR in the fiber feedthrough on the right of component (A). The section of the fiber inside the cryostat is represented by (B) and (C). (B) and (C) also represent the fiber curling shown in Fig.~\ref{fig:long_curling} and \ref{fig:fiber_heat_sink}, respectively. Finally, (D) corresponds to the TES optical stack. In every boundary between components, more BBR could couple to the fiber. However, this contribution is neglected given the lower temperature inside the cryostat.
    }        
 \label{fig:BBR_simulation_scheme}
\end{figure}

In this section, we present an example of the assembly of a specific simulation to recreate the conditions of the extrinsic background measurement discussed in this paper. Fig.~\ref{fig:BBR_simulation_scheme} shows a scheme of the components considered in this simulation.

The BBR from the experiment room at 295~K couples into the optical fiber, which has a numerical aperture $\mathrm{NA} = 0.14$ and a core radius $r_{\mathrm{core}} = 3.1~\mathrm{\upmu m}$ given by the manufacturer. The numerical aperture is related to $\theta_{\mathrm{max}}$ as $\mathrm{NA} = n_{\mathrm{air}}\sin{\theta_{\mathrm{max}}} \approx \sin{\theta_{\mathrm{max}}}$, where $n_{\mathrm{air}} = 1.0003$ is the index of refraction of the medium containing the optical fiber, in this case, the air. The integrals in Eq.~\ref{eq:BBR_components} can already be computed (Eq.~\ref{eq:area_solid_angle}) as the BBR term and the transmissions do not depend on the angle $\theta$ or the area $A_{\mathrm{core}}$ of the fiber's core, which is considered a circle:

\begin{equation}
\label{eq:area_solid_angle}
    \begin{split}
    \iint_{A, \Omega} \cos{\theta} \, d\Omega \, dA & = \\
    A_{\mathrm{core}}\Omega_{eff} & = \pi^2 (NA)^2 r^2_{\mathrm{core}} = 1.86~\mathrm{\upmu m^2}.
    \end{split}
\end{equation}

Afterwards, the photons travel through the optical fiber (A, in Fig.~\ref{fig:BBR_simulation_scheme}) in the warm. In this section, the absorbed photons in the optical fiber will be compensated for by the photons produced by the optical fiber. This is equivalent to the production of BBR just before the fiber enters the cryostat. 

\begin{figure}[!h]
    \centering
    \captionsetup[subfloat]{captionskip=5pt}
    \subfloat[]{
        \includegraphics[width=0.80\columnwidth]{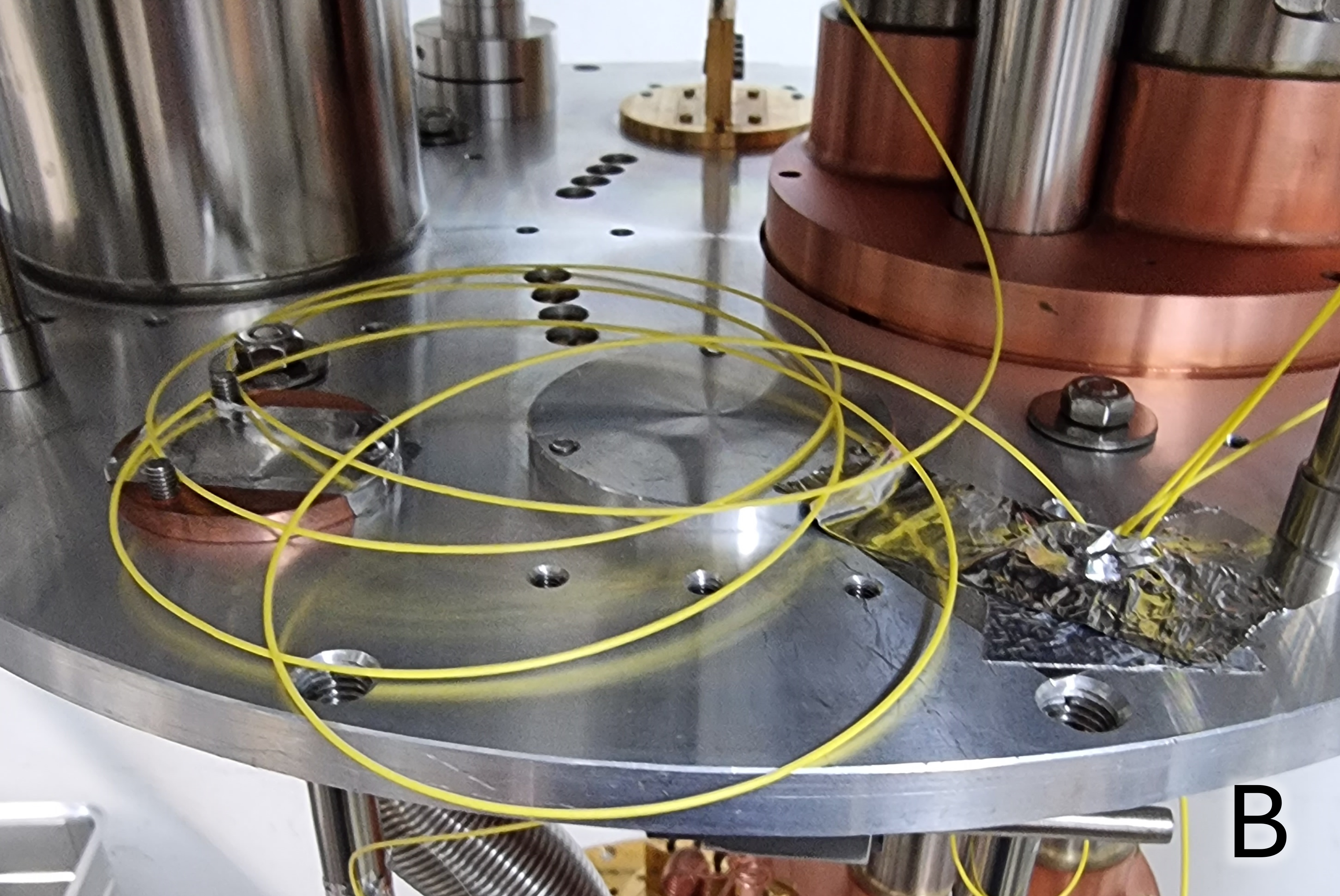}
        \label{fig:long_curling}
    }    
    
    \subfloat[]{
        \includegraphics[width=0.80\columnwidth]{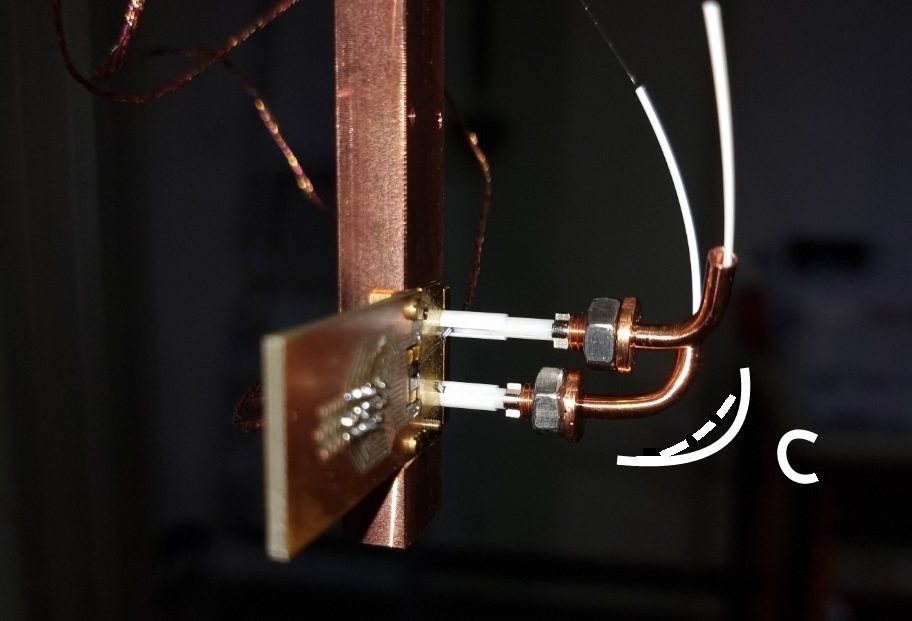}
        \label{fig:fiber_heat_sink}
    }
    \caption{
        (a) Fiber curled inside the cryostat approximated as 4~loops of diameter $d = 11.2~\mathrm{cm}$, component B in the simulation, with transmission $\eta_{\mathrm{curl1}}$, (b) Bending of the optical fiber due to the heat sink (copper parts covering the fiber ends), represented in the simulation as 1/4~loop of diameter $d = 2.5~\mathrm{cm}$, component C in the simulation. Note that, as the white curve shows, the piece is not described as a perfect arc of circumference as shown in the white dashed curve. This means that the curl diameter $d$ could be smaller, introducing one of the main uncertainties in Fig.~\ref{fig:simulation_ER_comparison}.
    }
    \label{fig:setup_curling}
\end{figure}

Once the fiber enters the cryostat, its transmission is considered $\eta_{\mathrm{fiber}}(E, \mathrm{4~m})$, with a length estimated at 4~m. Also, inside the cryostat, the fiber is curled, as shown in Fig.~\ref{fig:long_curling}, with four loops of 10~cm diameter, approximated to 11.2~cm since the curling data for 10~cm was not available. This is equivalent to the component B shown in Fig.~\ref{fig:BBR_simulation_scheme}, with transmission $\eta_{\mathrm{curl1}}(E, 4)$. Another curling is included to account for the bending of the fiber in its heat sink (Fig.~\ref{fig:fiber_heat_sink}) just before connecting it to the TES. Its shape is estimated to produce a quarter of a loop of diameter 2.5~cm, with transmission $\eta_{\mathrm{curl2}}(E, 1/4)$. Afterwards, the absorbance of the TES is added as the last optical component. Then, the rate of BBR $\frac{dN(E)}{dE}$ as a function of temperature and energy arriving at the TES is given by:
\begin{equation}
\label{eq:total_BBR_rate}
    \begin{split}
        \frac{d}{dE} N (E) & = \\
        A_{\mathrm{core}}\Omega_{eff} & \cdot B(E, T) \cdot \eta_{\mathrm{fiber}} \cdot \eta_{\mathrm{curl1}} \cdot \eta_{\mathrm{curl2}} \cdot \eta_{\mathrm{TES}}.
    \end{split}    
\end{equation}

\begin{figure}[!t]
    \centering
    \includegraphics[width=0.40\textwidth]{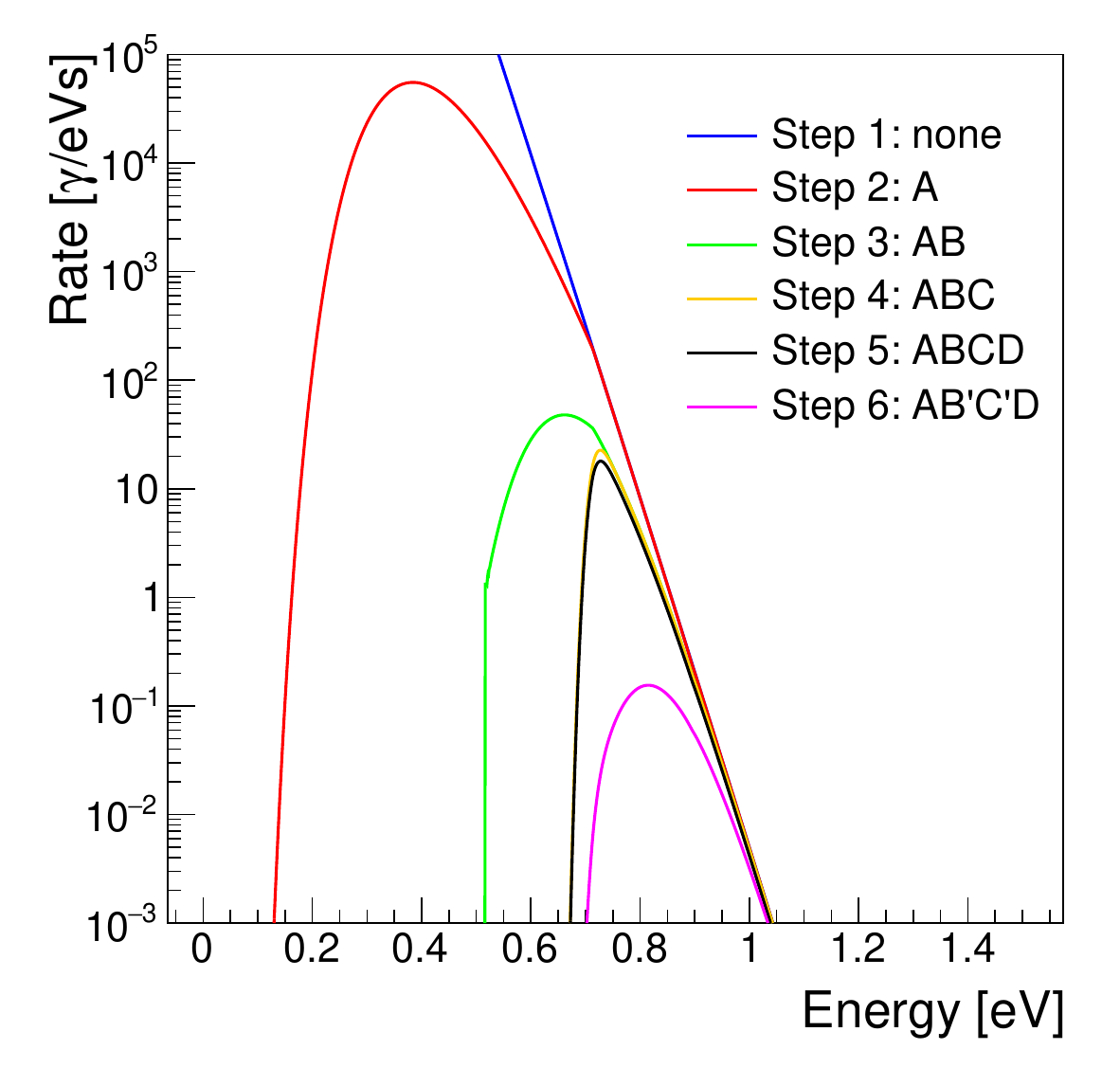}
    \vspace{-15pt}
    \caption{
        BBR spectrum computed from Eq.~\ref{eq:total_BBR_rate}, at a temperature of 295~K, step by step including optical components: Step 1 with no components, step 2 adding optical fiber transmission $\eta_{\mathrm{fiber}}$~(A), step 3 adding curling $\eta_{\mathrm{curl1}}$ with 4 loops with diameter $d = 11.2~\mathrm{cm}$~(B), step 4 adding curling $\eta_{\mathrm{curl2}}$ with 1/4 loops with diameter $d = 2.5~\mathrm{cm}$~(C), step 5 adding the reflectance of the TES optical stack~(D), step 6 adds alternative curling~(B'C') considered for the lower limit in Fig.~\ref{fig:simulation_ER_comparison}.
    }
    \label{fig:BBR_step_by_step}
\end{figure}

In Fig.~\ref{fig:BBR_step_by_step}, the effects of the individual optical components on the BBR spectrum computed with Eq.~\ref{eq:total_BBR_rate} are shown. The transmission of the optical fiber already cuts energies lower than 0.2~eV (step 2). The fiber curling reduces the rate of BBR at energies lower than 0.7~eV significantly (steps 3 and 4). Finally, as expected from section~\ref{sec:opt_components}, the influence of the absorbance of the TES optical stack is not noticeable in this case, as evidenced in Fig.~\ref{fig:BBR_step_by_step}, step 5.

\begin{figure*}[!tbp]
    \centering
        \includegraphics[width=0.8\textwidth]{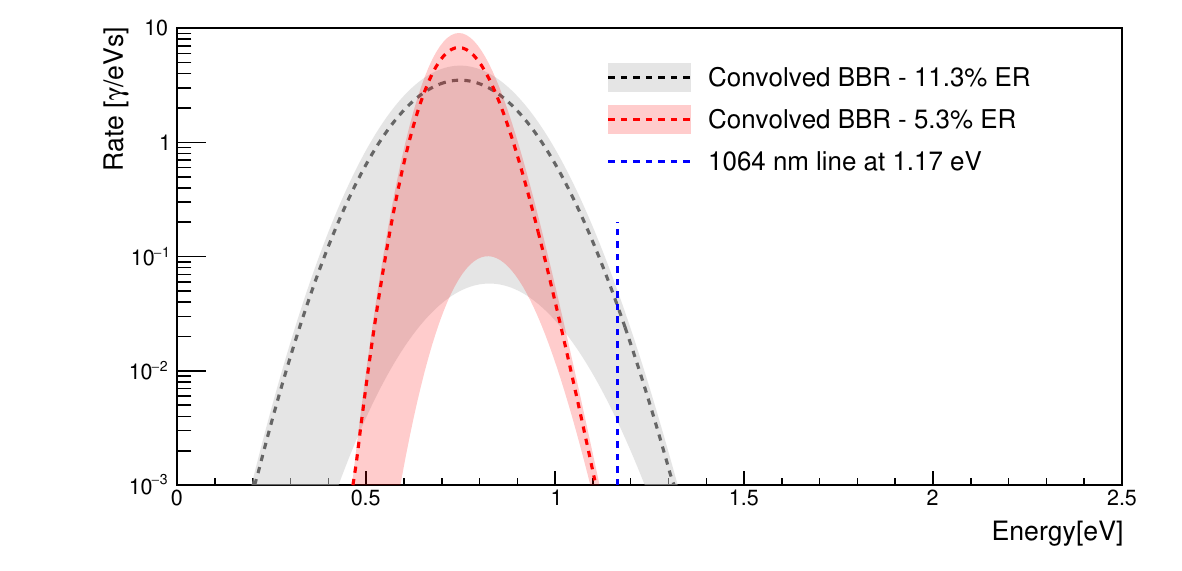}
        \vspace{-10pt}        
    \caption{
        The dashed curves represent the BBR spectra simulated at a temperature of 295~K, as explained in section~\ref{sec:example}, convoluted with different TES energy resolutions (11.3~\% in gray, 5.3~\% in red). The line corresponding to the energy of 1064~nm photons is shown in blue for reference. The upper limit of the bands is produced by varying the temperature by 2~K. The lower bound is produced by the assumption of more attenuation due to the non-accounted bending of the fiber inside the cryostat.
    }        
 \label{fig:simulation_ER_comparison}
\end{figure*}

The accuracy of Eq.~\ref{eq:total_BBR_rate} is affected mainly by the fiber curling component $\eta_{\mathrm{curl}}$ and the temperature used to compute $B(E, T)$.
As we do not expect other potential sources of uncertainty to contribute significantly to the total uncertainty compared to the effects of temperature variations and fiber curling, they were neglected.
Small variations in $T$ or the radius of the curling can produce changes by one order of magnitude in the results of the expected photon rate $d N (E)/dE$. This can be seen in Fig.~\ref{fig:simulation_ER_comparison}, where the upper values in the uncertainty bands are produced by a variation in the room's temperature by 2~K. The largest variation, represented in the lower values, originates by adding a 1/2-loop of 4.2~cm diameter and assuming a 1/4-loop of 1.2~cm diameter instead of 2.5~cm in $\eta_{\mathrm{curl2}}$ (extra step represented as step 6 in Fig.~\ref{fig:BBR_step_by_step}). The 4.2~cm half-loop could result from a non-accounted bending in another region of the cryostat since the fiber is not installed in a straight line inside the cryostat but rather needs to curve around the inner components. The sharper quarter-loop of 1.2~cm diameter instead of 2.5~cm in $\eta_{\mathrm{curl2}}$ is motivated 
by the deviation of the heat sink's arc from a perfect circle shown in Fig.~\ref{fig:fiber_heat_sink}.

The distributions shown in steps 5 and 6 of Fig.~\ref{fig:BBR_step_by_step} are then folded with the energy resolution and the time resolution using Eq.~\ref{eq:total_folded_BBR_rate}. Taking into account the energy resolution of 11.3~\% for 1064~nm photons, i.e. $\sigma_{\mathrm{1064nm}} = 0.13~\mathrm{eV}$, obtained from the data analysis described in~\cite{rubiera_gimeno_tes_2023}, the expected photon rate on the TES shown in gray in Fig.~\ref{fig:simulation_ER_comparison} is obtained.
A contribution typical for pile-up photons is negligible as it is highly suppressed by the fiber curling.

The main interest for ALPS~II lies in the reduction of the number of background events in the region surrounding the energy of 1064~nm photons. Therefore, considering both the TES energy resolution and the distribution of 1064~nm photons in the energy spectrum as a Gaussian, regions of $1\sigma$, $2\sigma$, $3\sigma$ are proposed, with $\sigma = \sigma_{\mathrm{1064nm}} = 0.13~\mathrm{eV}$.
The analysis efficiency in simulation is then defined as the efficiency resulting from excluding 1064\,nm photons outside the $\sigma$ ranges.
Table~\ref{tab:BBR_sim_comparison} presents the simulated spectra results and computed BBR rates with 11.3~\% energy resolution, with the upper and lower limits of the uncertainty bands.
The rates are shown for several $\sigma$ regions around 1064~nm photon energy.
\begin{table}[!h]
    \centering
    \caption{Rates for simulated BBR with 11.3~\% energy resolution.}
    \begin{tabular}{|c|c|c|c|}
    \hline
        Range ($\sigma$) &
        \makecell{Analysis \\ efficiency} & 
        \makecell{Simulated \\ lower rate} & 
        \makecell{Simulated \\ upper rate} \\
    \hline
        $-1, 1$     & $68.2\%$  & $1.5 \cdot 10^{-3}~\mathrm{cps}$ & $3.1 \cdot 10^{-2}~\mathrm{cps}$ \\
    \hline
        $-2, 2$     & $95.4\%$  & $6.2 \cdot 10^{-3}~\mathrm{cps}$ & $2.1 \cdot 10^{-1}~\mathrm{cps}$ \\
    \hline
        $-3, 3$     & $99.7\%$  & $1.3 \cdot 10^{-2}~\mathrm{cps}$ & $7.0 \cdot 10^{-1}~\mathrm{cps}$ \\
    \hline
        $~0, 3$     & $49.8\%$  & $2.0 \cdot 10^{-4}~\mathrm{cps}$ & $2.2 \cdot 10^{-3}~\mathrm{cps}$ \\
    \hline
        $-1, 3$     & $84.0\%$  & $1.5 \cdot 10^{-3}~\mathrm{cps}$ & $3.1 \cdot 10^{-2}~\mathrm{cps}$ \\
    \hline
    \end{tabular}    
    \label{tab:BBR_sim_comparison}
\end{table}

Given the asymmetry of the BBR spectrum, regions given by $[-\sigma, 3\sigma]$ or $[0, 3\sigma]$ could be a good trade-off to reduce the background while maintaining high analysis efficiency, but even in the lower rate column in Table~\ref{tab:BBR_sim_comparison}, the predicted rates of $r = 1.5 \cdot 10^{-3}~\mathrm{cps}$ and $r = 2.0 \cdot 10^{-4}~\mathrm{cps}$ are still well above the requirements for ALPS~II of $r < 7.7 \cdot 10^{-6}~\mathrm{cps}$~\cite{epsrikhav}.

To reduce these rates, the use of an optical filter, as the one shown in Fig.~\ref{fig:simulation_optical_filter}, in the cold region of the cryostat has been proposed. However, this alternative suffers from misalignment and the shifting of the filtered wavelength due to the cold temperatures, which reduce the transmission efficiency for the signal at 1064~nm and, consequently, the signal-to-noise ratio. Strategies to overcome this are under investigation. To study the effects of the filter in the BBR spectrum, it has been implemented in the simulation framework\footnote{The optical filter was not included in the simulation example in section~\ref{sec:example}.}.
In addition to the optical filter, the use of more loops in the fiber curling or lower radii could help to reduce the background, but this strategy has limits, as a low radius of curling can also reduce the efficiency for 1064~nm photons.

Another proposed solution is to improve the TES's energy resolution, which in this paper is achieved through data analysis. As shown in Fig.~\ref{fig:simulation_ER_comparison}, a better energy resolution significantly reduces the background due to BBR in the region surrounding 1.17~eV (1064~nm photons). This motivated the analysis improvements originally developed in~\cite{RubieraGimeno_eps} and applied here, resulting in an energy resolution of 5.3~\%, with $\sigma_{1064\,\mathrm{nm}} = 0.061~\mathrm{eV}$. A short description of the data analysis is done in section~\ref{sec:extrinsiccompartison}.
Taking into account the improved energy resolution, the spectrum shown in red in Fig.~\ref{fig:simulation_ER_comparison} is obtained. The rates computed for simulated BBR with 5.3~\% energy resolution, with the upper and lower limits of the uncertainty bands, are shown in Table~\ref{tab:BBR_sim_comparison_better}.

\vspace{-10pt}
\begin{table}[H]
    \centering
    \caption{
        Rates for simulated BBR with 5.3~\% energy resolution.
    }
    \begin{tabular}{|c|c|c|c|}
    \hline
        Range ($\sigma$) &
        \makecell{Analysis \\ efficiency} & 
        \makecell{Simulated \\ lower rate} & 
        \makecell{Simulated \\ upper rate} \\
    \hline
        $-1, 1$     & $68.2\%$  & $2.2 \cdot 10^{-5}~\mathrm{cps}$ & $4.6 \cdot 10^{-5}~\mathrm{cps}$ \\
    \hline
        $-2, 2$     & $95.4\%$  & $1.5 \cdot 10^{-4}~\mathrm{cps}$ & $4.1 \cdot 10^{-4}~\mathrm{cps}$ \\
    \hline
        $-3, 3$     & $99.7\%$  & $7.7 \cdot 10^{-4}~\mathrm{cps}$ & $3.3 \cdot 10^{-3}~\mathrm{cps}$ \\
    \hline
        $~0, 3$     & $49.8\%$  & $0.3 \cdot 10^{-5}~\mathrm{cps}$ & $0.5 \cdot 10^{-5}~\mathrm{cps}$ \\
    \hline
        $-1, 3$     & $84.0\%$  & $2.3 \cdot 10^{-5}~\mathrm{cps}$ & $4.6 \cdot 10^{-5}~\mathrm{cps}$ \\
    \hline
    \end{tabular}
    \label{tab:BBR_sim_comparison_better}
\end{table}
The analysis efficiency is the same as defined before and the rates are shown for several $\sigma$ regions around 1064~nm photon energy. 

The new background rates in the regions given by $[-\sigma, 3\sigma]$ or $[0, 3\sigma]$ are predicted in the upper limit as $4.6 \cdot 10^{-5}~\mathrm{cps}$ and $ 5 \cdot 10^{-6}~\mathrm{cps}$, respectively, the latter already fulfilling the limit of the ALPS~II requirements. If the measured data were to follow the distribution in Fig.~\ref{fig:simulation_ER_comparison}, the integration regions could be optimized to maximize significance taking also the efficiency into account, as done in Ref.~\cite{manuel_ML}.

\section{Comparison with extrinsic background measurement}
\label{sec:extrinsiccompartison}

In order to evaluate the contribution of BBR, an extrinsic background dataset was measured for 72~hours by covering the warm end of the optical fiber that goes to the cryostat with a black cloth. The lab was dark and at approximately $295~\mathrm{K}$. The fiber was curled inside the cryostat and connected to the TES. The conditions were similar to those depicted in the simulation described in section~\ref{sec:example}, although the precise path of the fiber was not fully documented, which could have led to unmonitored bending. A 1064~nm sample was taken before the start of the measurement for reference. For data taking, a trigger threshold was set at 20~mV for both measurements as done in~\cite{rubiera_gimeno_tes_2023}, and the measurements were processed with a fitting-based analysis. The fitting was performed on the signal in the time and frequency domains. This is described in~\cite{jltprikhav} and \cite{RubieraGimeno_eps}, respectively. The pulses in the time domain are fitted to the following equation obtained from a phenomenological approach:
\begin{equation}
\label{eq:phenomenological}
	\begin{gathered}
        U_{ph}(t) = \qquad \qquad \qquad \qquad \quad\\
        		- \frac{2A_{ph}}{
			\exp{ \left \{ 
			 {-\frac{1}{\tau_{\mathrm{rise}}}(t-t_0)}
			\right \} } 
			+ \exp{ \left \{ 
			 {\frac{1}{\tau_{\mathrm{decay}}}(t-t_0)}
			\right \} }} + V_0,
    \end{gathered}
\end{equation}
where the parameters $A_{ph}$, $\tau_{\mathrm{rise}}$, and $\tau_{\mathrm{decay}}$ define the shape of the pulse and are related to its amplitude, rise time, and decay time, respectively. The parameters $t_0$ and $V_0$ correspond to the position of the pulse in the time window and the voltage offset.

For the fitting in the frequency domain, the Fourier transformation of the pulses is computed. The same is done to the function that predicts the pulse shape, obtained from the small signal theory for the TES signal~\cite{Irwin2005, RubieraGimeno_eps}. The function used for fitting in the frequency domain is described by the following equation:
\begin{equation}
\label{eq:FFT}
\begin{gathered}
\mathscr{F}[U_{SST}(t)](f) = 
    -A_{SST}(\tau_--\tau_+) \exp{\{-2\pi f i t_0\}} \times \\
    \qquad \qquad \qquad \times \frac{[1-(2\pi f)^2\tau_+\tau_-] - i2\pi f(\tau_++\tau_-)}
    {[1+\tau_+^2(2\pi f)^2][1+\tau_-^2(2\pi f)^2]}
\end{gathered}
\end{equation}

\begin{figure*}[!t]
    \centering
    \captionsetup[subfloat]{captionskip=-12pt}
    \subfloat[]{
        \includegraphics[width=0.7\columnwidth]{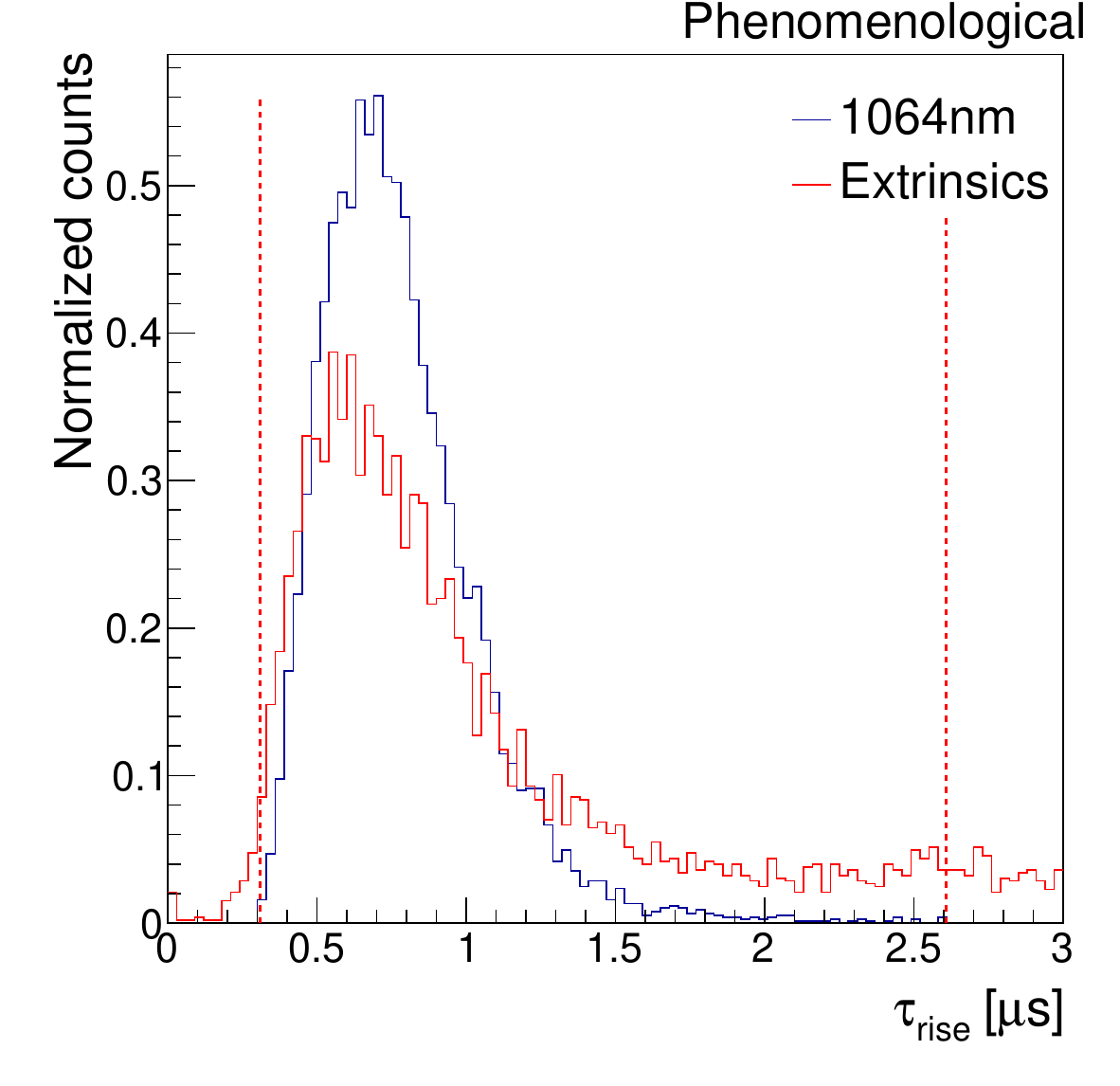}
        \label{fig:rise_cut_time}
        }
    \hfil
    \subfloat[]{
        \includegraphics[width=0.7\columnwidth]{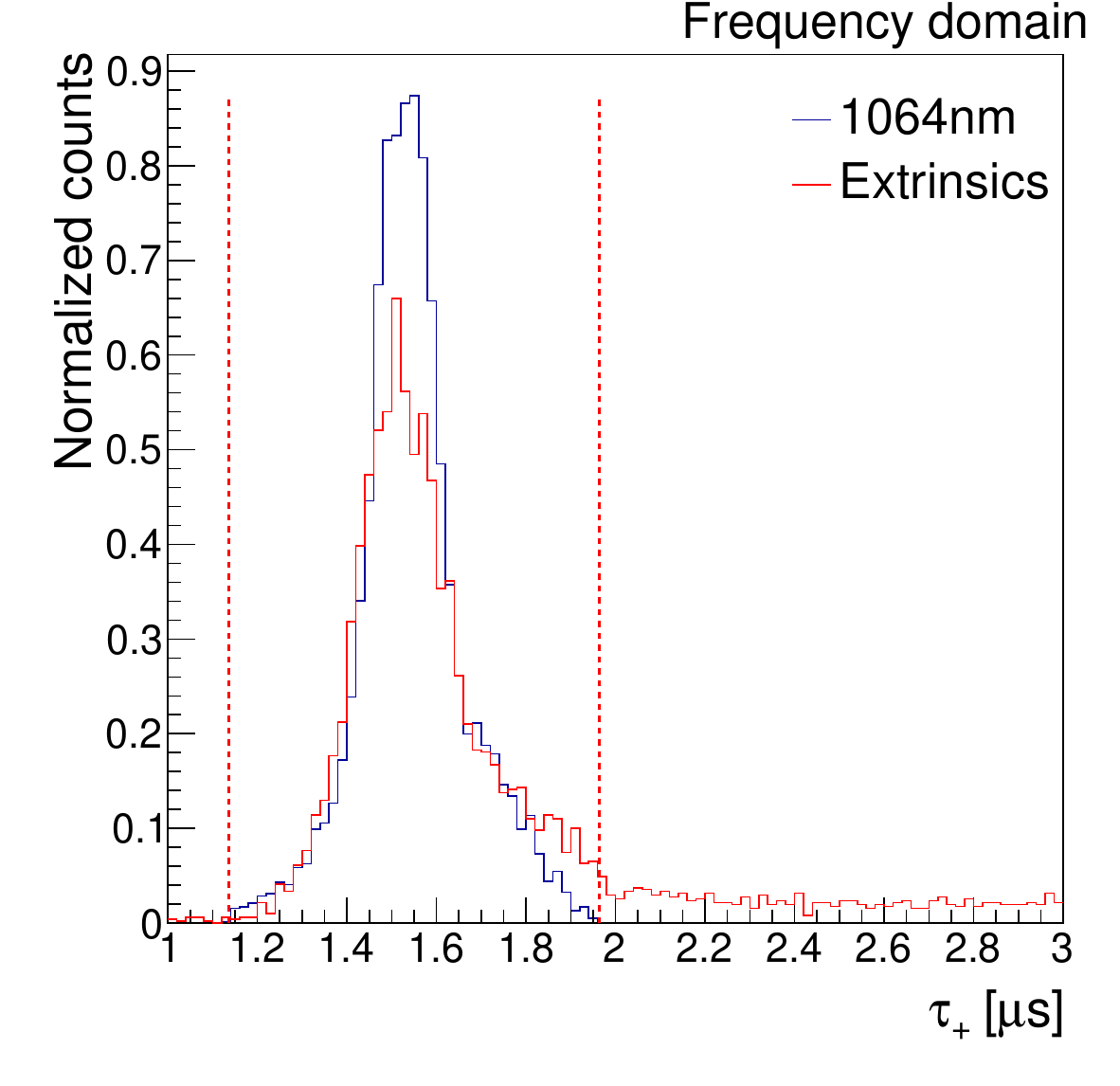}
        \label{fig:rise_cut_frequency}
        }
    \\
    \subfloat[]{
        \includegraphics[width=0.7\columnwidth]{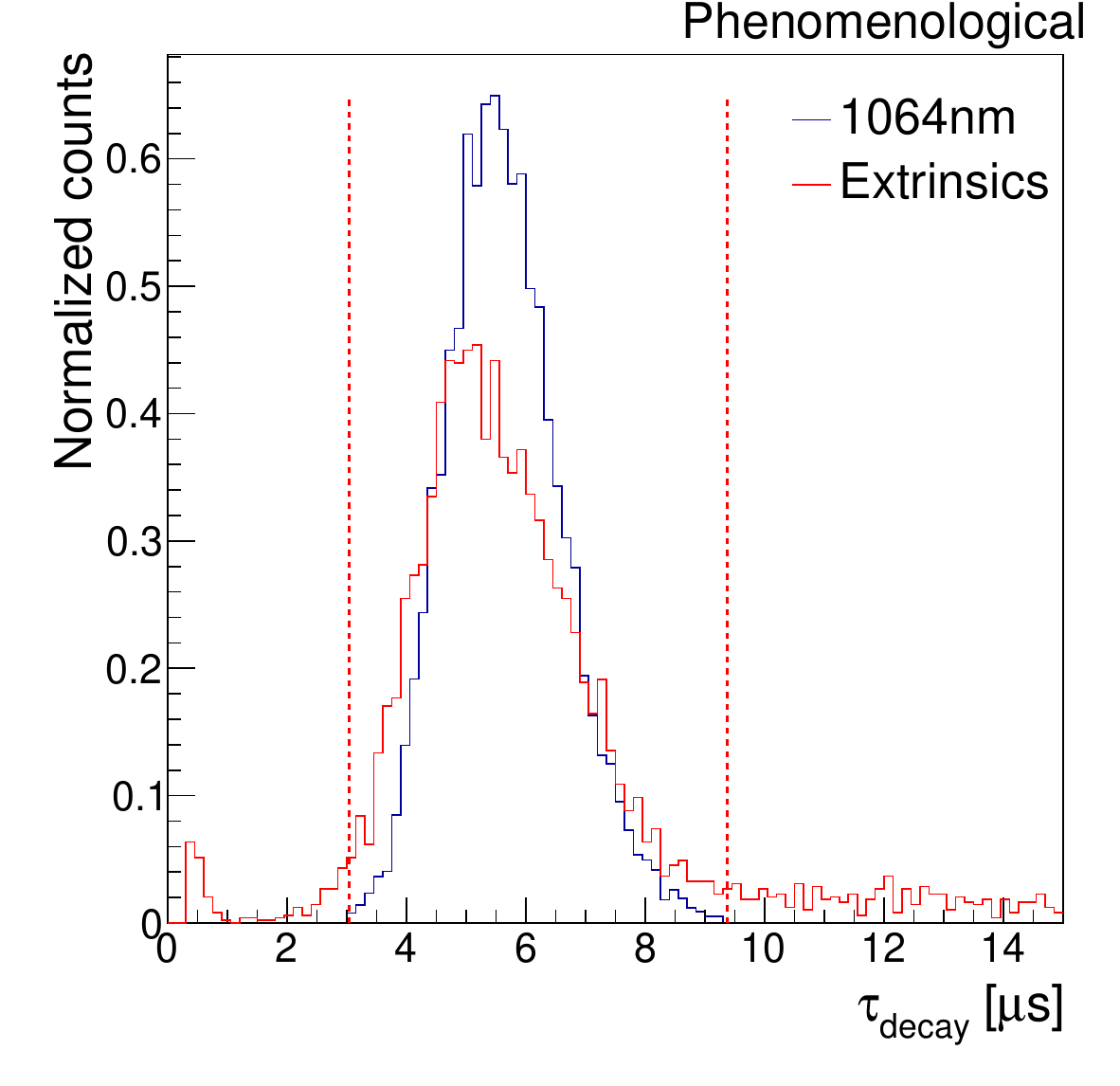}
        \label{fig:decay_cut_time}        
        }
    \hfil
    \subfloat[]{
        \includegraphics[width=0.7\columnwidth]{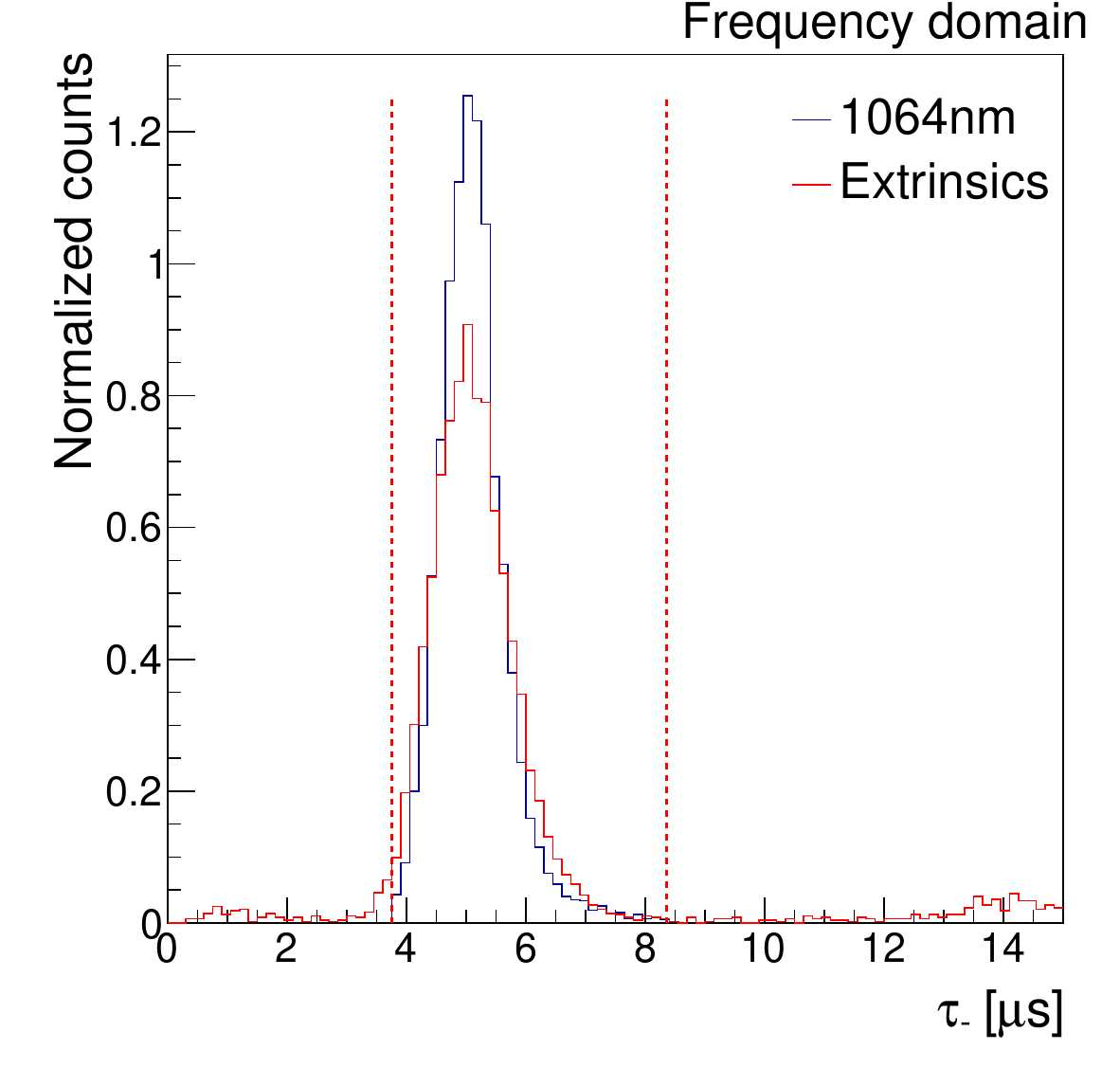}
        \label{fig:decay_cut_frequency}
        }
    \\
    \subfloat[]{
        \includegraphics[width=0.7\columnwidth]{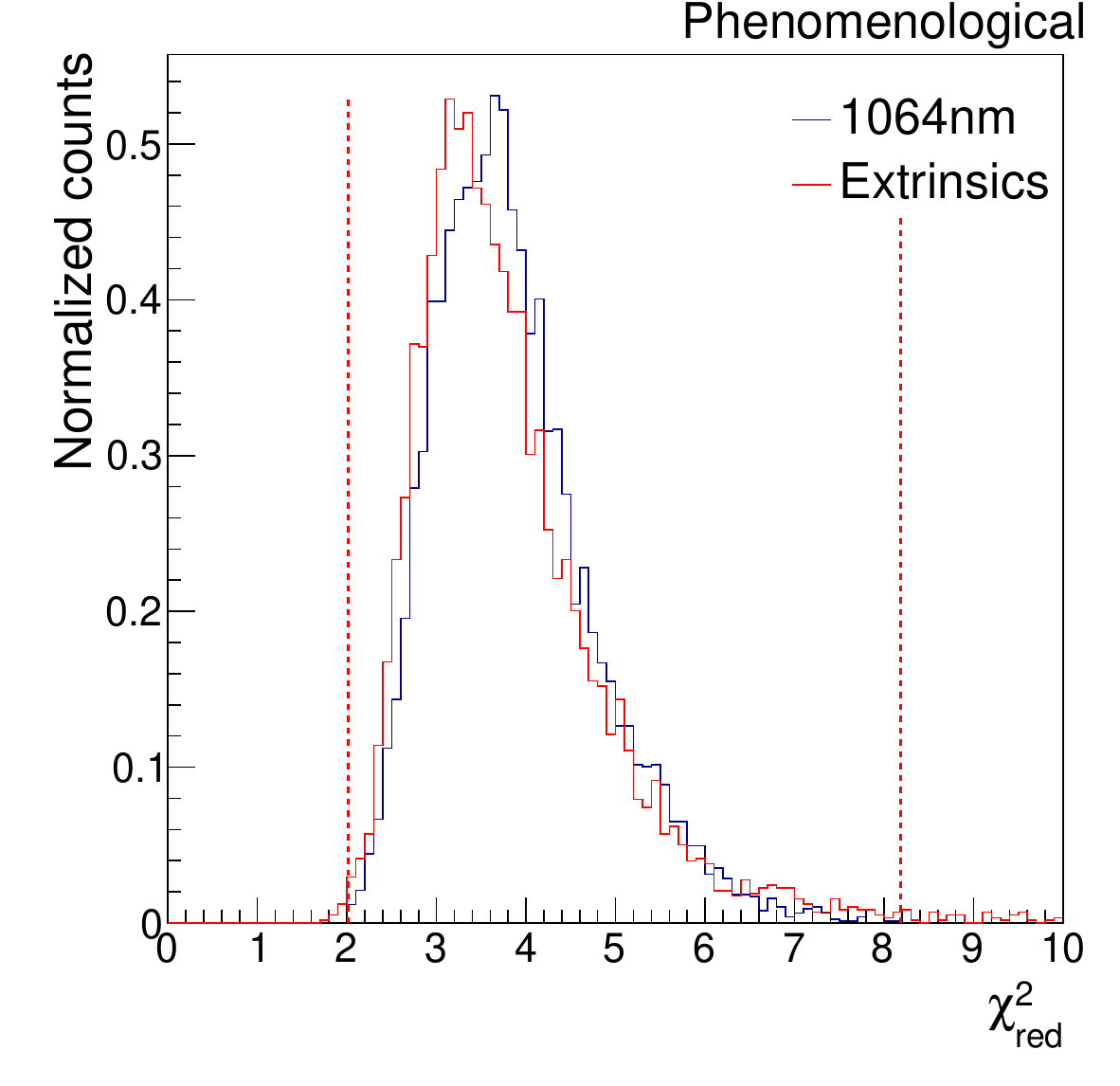}
        \label{fig:chi_red_cut_time}        
        }
    \hfil
    \subfloat[]{
        \includegraphics[width=0.7\columnwidth]{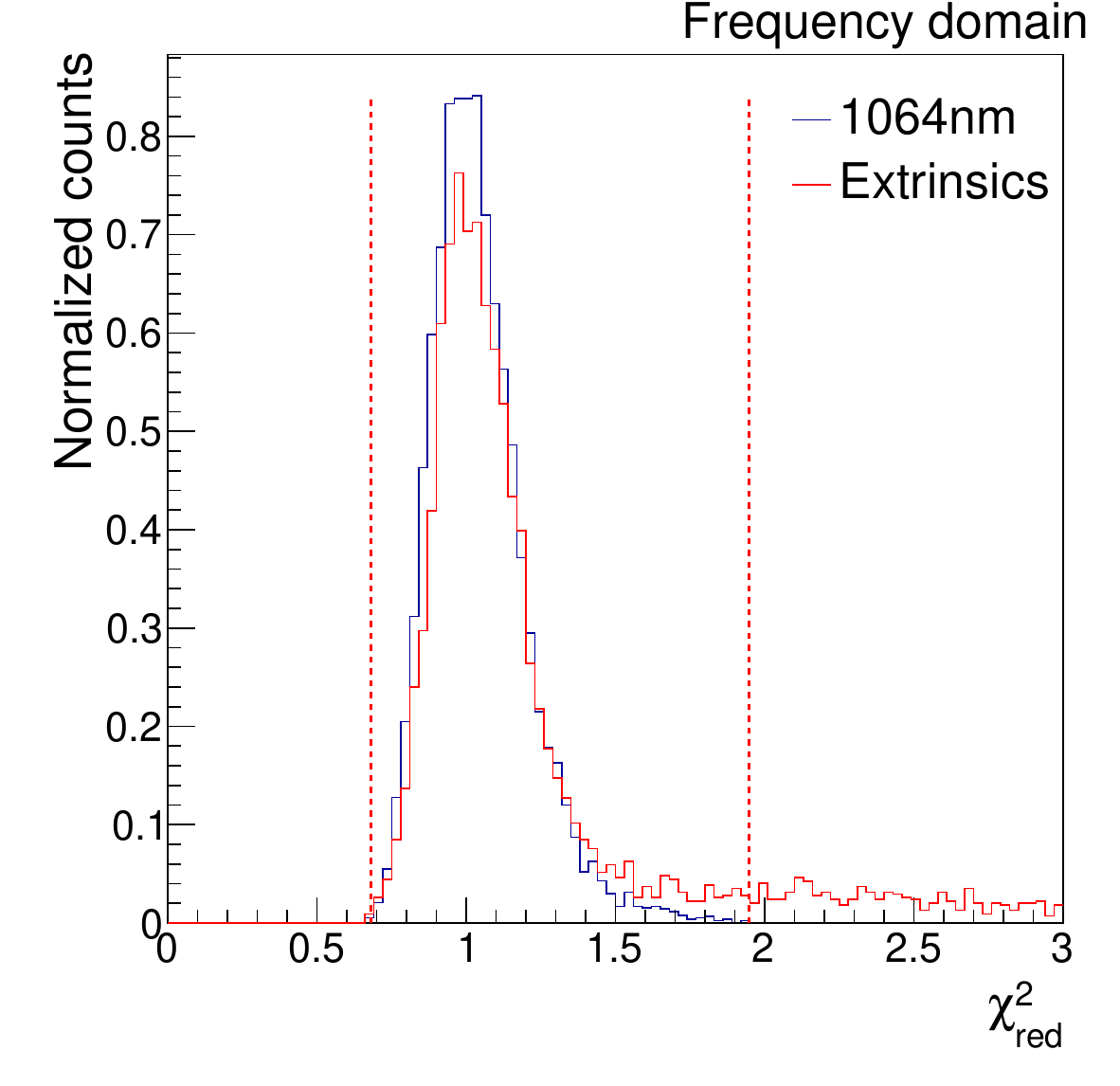}
        \label{fig:chi_red_cut_frequency}
        }
    \caption{
        Distribution of parameters obtained by fitting the phenomenological model in a, c and e, and the frequency domain SST model in b, d and f, to measured 1064~nm pulses and pulses from an extrinsics measurement. The cuts (dashed red lines) exclude the pulses that lie outside the $[-3\sigma, 3\sigma]$ region of the parameters: (a) $\tau_{\mathrm{rise}}$, (c) $\tau_{\mathrm{decay}}$ and (e) $\chi^2_{red}$ in the time domain, and (b) $\tau_{+}$, (c) $\tau_{-}$ and (e) $\chi^2_{red}$ in the frequency domain.
    }
    \label{fig:par_dist_cuts}
\end{figure*}

\begin{figure*}[!t]
    \centering
        \includegraphics[width=0.8\textwidth]{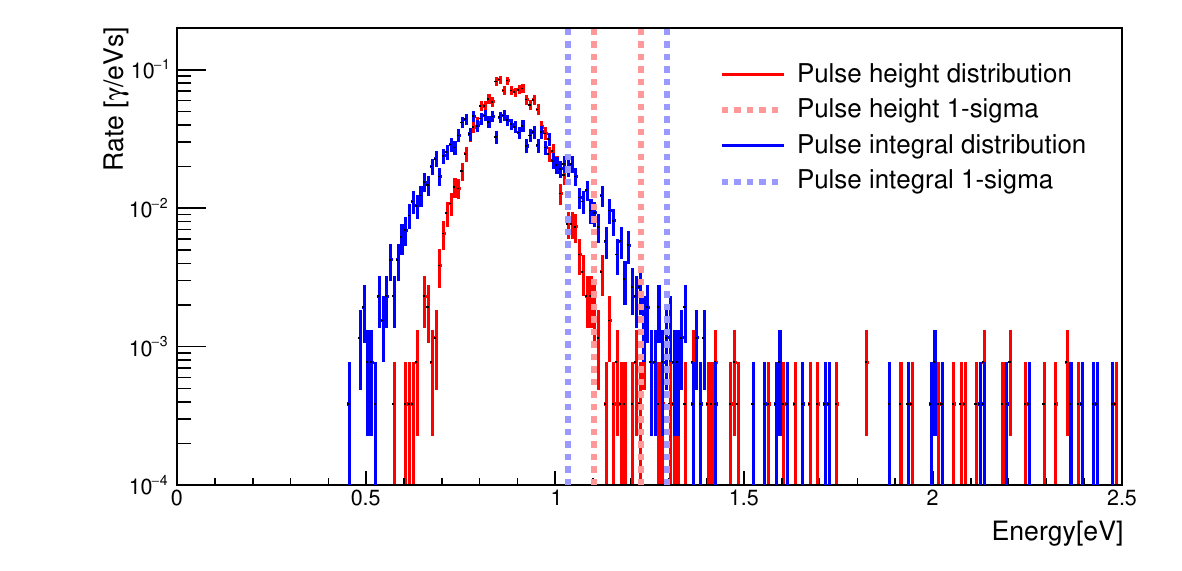}
        \vspace{-10pt}        
    \caption{
        Measured extrinsics energy spectrum computed using the pulse integral from the time domain analysis (blue) and the pulse height from the frequency domain analysis (red).
    }        
 \label{fig:ER_integral_height}
\end{figure*}

The parameters $A_{SST}$, $\tau_{+}$, $\tau_{-}$, and $t_0$ are analogous to the parameters $A_{ph}$, $\tau_{\mathrm{rise}}$, $\tau_{\mathrm{decay}}$ and $t_0$ in the phenomenological model.

Analogously to~\cite{epsrikhav}, cuts were imposed on the rise and decay times of the pulses and the reduced $\chi^2$ from the fitting procedure. These parameters were chosen given that they should not depend on the energy of the arriving photon according to TES theory~\cite{Irwin2005}. The cuts were performed on the interval $[-3\sigma, 3\sigma]$ based on the 1064\,nm calibration pulses to keep the acceptance of 1064\,nm photons high.
In the left column of Fig.~\ref{fig:par_dist_cuts}, the distributions of rise time, decay time and reduced $\chi^2$ for 1064\,nm photons from the fitting analysis in the time domain and the corresponding cuts can be seen. The cuts imposed on the rise and decay times of the pulses and the reduced $\chi^2$ from the fitting procedure based on the frequency domain are shown in the right column of Fig.~\ref{fig:par_dist_cuts}.
The overall acceptance is then 98.4~\% for the cuts in the six distributions of 1064\,nm photon parameters. The cuts are applied to the results of the analysis of the extrinsics dataset, resulting in a reduction of the background rate from $4.4 \cdot 10^{-2}~\mathrm{cps}$ to $1.5 \cdot 10^{-2}~\mathrm{cps}$. This indicates that the surviving events are mostly photons, which is consistent with the optical fiber being connected to the TES. In contrast, in~\cite{epsrikhav}, the optical fiber was disconnected from the TES, and the cuts applied in the measured data reduced the background rate by several orders of magnitude down to $6.9 \cdot 10^{-6}~\mathrm{cps}$.

The integral and the pulse height of the photon pulses are assumed to be proportional to the photon's energy, equivalent to a linear response of the TES. A calibration is performed with the integral $I_{1064\mathrm{nm}}$ and the pulse height $h_{1064\mathrm{nm}}$ of the 1064~nm photons, where $I_{1064\mathrm{nm}}$ and $h_{1064\mathrm{nm}}$ correspond to the center of the Gaussian functions that are fitted to the respective 1064~nm distributions in measured data. Then, the measured energy of the photon is assumed to be $E = E_{1064\mathrm{nm}}I/I_{1064\mathrm{nm}}$ and $E = E_{1064\mathrm{nm}}h/h_{1064\mathrm{nm}}$. Both values are chosen to compare the BBR rates before and after the improvement of the energy resolution shown in \cite{RubieraGimeno_eps}. The events surviving the aforementioned cuts are accumulated in a histogram. They are scaled, dividing them by the measurement's total time and the energy bin's size. The results using both magnitudes, pulse integral and pulse height, are shown in Fig.~\ref{fig:ER_integral_height}. 

These results match the predictions from the simulation about the shape of the BBR spectrum in Fig.~\ref{fig:simulation_ER_comparison}.
Improving energy resolution narrows the BBR spectrum and reduces the 1064~nm photon signal region, effectively reducing the extrinsic background rates in the 1064~nm region. The measured rates with the optimized analysis are summarized in Table~\ref{tab:BBR_energy_res_better} for the same $\sigma$ ranges defined before.  Table~\ref{tab:BBR_energy_res_better} shows a comparison between the rates computed for measured extrinsics data using the pulse integral from the time domain analysis ($I_{Ph}$) and the pulse height from the frequency domain analysis ($h_{FFT}$) to calculate the energy of the photons.
\begin{table}[h!]
    \centering
    \caption{
    Comparison of the rates for measured extrinsics data using two different methods to compute the energy.
    }
    \label{tab:BBR_energy_res_better}
    \begin{tabular}{|c|c|c|c|}
    \hline
        Range ($\sigma$) & Efficiency & Rate ($I_{Ph}$) & Rate ($h_{FFT}$) \\
    \hline
        $-1, 1$     & $60.5\%$ &    $1.7 \cdot 10^{-3}~\mathrm{cps}$     & $1.2 \cdot 10^{-4}~\mathrm{cps}$ \\
    \hline
        $-2, 2$     & $84.5\%$ &    $5.6 \cdot 10^{-3}~\mathrm{cps}$     & $4.1 \cdot 10^{-4}~\mathrm{cps}$ \\
    \hline
        $-3, 3$     & $88.3\%$ &    $1.1 \cdot 10^{-2}~\mathrm{cps}$     & $1.5 \cdot 10^{-3}~\mathrm{cps}$ \\
    \hline
        $~0, 3$     & $44.2\%$ &    $4.2 \cdot 10^{-4}~\mathrm{cps}$     & $6.9 \cdot 10^{-5}~\mathrm{cps}$ \\
    \hline
        $-1, 3$     & $74.3\%$ &    $1.9 \cdot 10^{-3}~\mathrm{cps}$     & $1.6 \cdot 10^{-4}~\mathrm{cps}$\\
    \hline
    \end{tabular}
\end{table}

The efficiency in Table~\ref{tab:BBR_energy_res_better} combines a cut acceptance of 98.4~\%, the efficiency resulting from excluding 1064~nm photons outside the $\sigma$ ranges, and the system detection efficiency from~\cite{RubieraGimeno_eps}. The system detection efficiency refers to the efficiency of the coupling of the optical fiber to the TES, with a current estimated value of 90~\%.

Overall, the improvement in energy resolution by a factor of 2 shown in~\cite{RubieraGimeno_eps} reduces the background rate in every range by an order of magnitude compared to the previous analysis. However, the measured rates of $1.6 \cdot 10^{-4}~\mathrm{cps}$ and $6.9 \cdot 10^{-5}~\mathrm{cps}$ in the regions given by $[-\sigma, 3\sigma]$ or $[0, 3\sigma]$ respectively, still do not satisfy the ALPS~II requirements of $r < 7.7 \cdot 10^{-6}~\mathrm{cps}$. The $[0, 3\sigma]$ region analysis efficiency is also below the ALPS~II efficiency requirement of 50~\%.

\begin{figure*}[t!]
    \centering
        \includegraphics[width=0.8\textwidth]{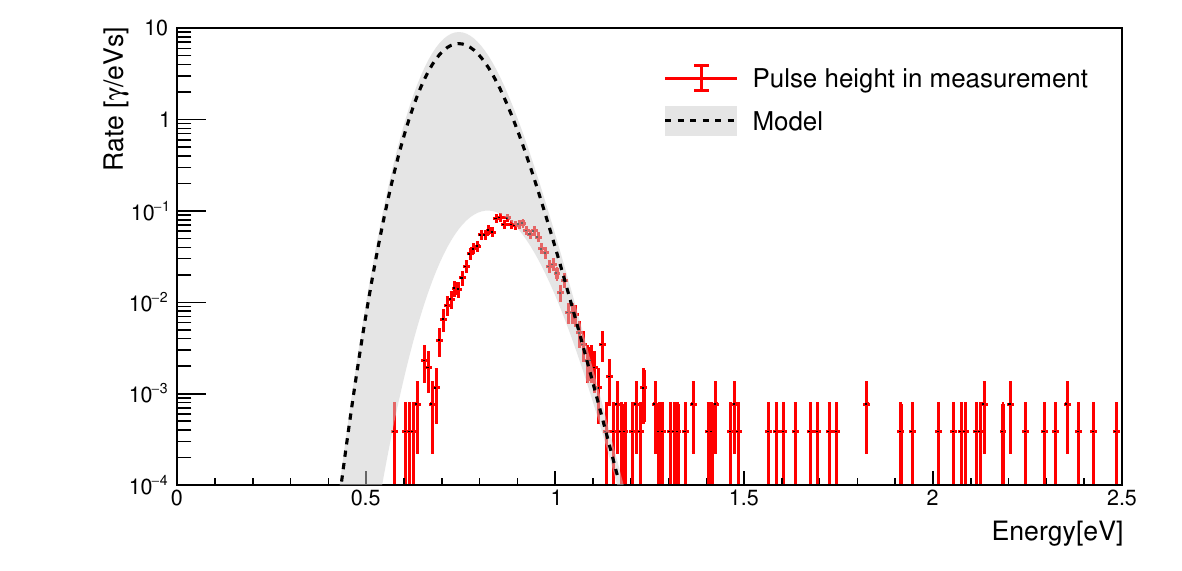}
        \vspace{-15pt}
    \caption{
        Simulated spectrum in Fig.~\ref{fig:simulation_ER_comparison}, folded with an energy resolution of 5.3~\%, compared to the measured rates from data. The pulse height computed from the analysis was scaled to energy using a 1064~nm sample as a reference. While the model accurately represents the data in the main region of interest, it does not effectively describe the low-energy part of the spectrum, which is susceptible to uncertainties introduced by fiber curling.
    }        
 \label{fig:BBR_height_comparison}
\end{figure*}

As shown in Fig.~\ref{fig:BBR_height_comparison}, the shape of the simulated spectrum is very similar to the one measured in the extrinsics measurement. This indicates the consistency of the measured background with BBR, although the uncertainties in the simulation limit its ability to fully describe the background. In particular, features such as the low energy edge and the distribution maximum are highly sensitive to the radius and the number of loops in the fiber curling. To improve the accuracy of the BBR simulations, detailed tracking of the path of the fiber inside the cryostat should be done in future measurements, as the curling has been identified as the main contributing factor (shown in the transition from step 5 to step 6 in Fig.~\ref{fig:BBR_step_by_step}), especially for low curling diameters. Furthermore, for measurements similar to the one analyzed in this section, a temperature sensor should be placed near the closed end of the fiber to monitor fluctuations during the measurement.

The simulation resembles the data trend in the region between 0.95~eV and 1.2~eV. Above 1.2~eV, several background events are present, possibly non-photon events surviving the cut discrimination. The cause of the excess of events in the 1-$\sigma$ signal region (between 1.10~eV and 1.22~eV), compared to the simulation, will be investigated, as the excess indicates that these events are not due to BBR.


\section{Conclusion and Outlook}
\label{sec:conclusion}

ALPS~II is expected to probe $g_{a\gamma \gamma}$ in a model-independent approach beyond present limits from astrophysics. Data collection is underway utilizing a heterodyne-based detection scheme. In parallel, a TES is being studied for a single photon detection scheme to complement the current measurements with the HET detection system. This would be particularly important in case an ALP signal is observed. Towards these scenarios, the TES background is being carefully studied and efforts are ongoing for its reduction.

In this paper, it has been shown that the primary background contributor in the extrinsics measurement setup is consistent with Black Body Radiation (BBR), aligning with predictions from previous studies~\cite{miller2007superconducting}. 
A framework has been developed to simulate the propagation of BBR to the TES, with adaptable features for various experimental conditions, with the ultimate aim of reducing this background source.
The simulation studies show that the low energy region of the BBR spectrum is filtered mainly by fiber curling. In addition, fiber curling allows the suppression of pile-up events in the BBR spectrum which could mimic 1064~nm photons.

Another important conclusion from the BBR simulation is that the background can be reduced by improving the energy resolution. This was achieved with the optimization of the data analysis described in~\cite{RubieraGimeno_eps}.
While the current measured background rate has not yet reached the ALPS II requirements, the measured rate of extrinsic events was reduced by an order of magnitude as a result of the improvement in energy resolution, in agreement with the BBR simulation predictions.
Additionally, the BBR simulation indicates a considerable reduction in the background rate by lowering the temperature in the environment around the optical fiber, e.g. decreasing the temperature of the lab to 10\textdegree C (283\,K) in the described setup would reduce the BBR background rate by a factor of 5.
This possibility merits further investigation.
The use of an optical filter inside the cryostat is also being explored as another approach for BBR background reduction.
A comparison of the measured spectrum with that predicted by the BBR simulation suggests, although statistically limited, the presence of non-Black Body events in the region around 1064~nm, which will require further study.


\begin{acknowledgments}
We thank Adriana Lita from the National Institute of Standards and Technology, USA, for the TES devices, and Joern Beyer from Physikalisch-Technische Bundesanstalt, Berlin, Germany, for the SQUID modules, vital advice and support. We thank Erika Garutti for helpful discussions. We thank our ALPS collaborators. We thank Harry Raymond for his assistance with the experimental measurements of fiber curling. Manuel Meyer acknowledges the European Research Council (ERC) support under the European Union's Horizon 2020 research and innovation program Grant agreement No. 948689 (AxionDM). We acknowledge the support by the Deutsche Forschungsgemeinschaft (DFG, German Research Foundation) under Germany’s Excellence Strategy – EXC 2121 ``Quantum Universe" - 390833306.
\end{acknowledgments}

\bibliography{simulation}

\begin{thebibliography}{26}%
\makeatletter
\providecommand \@ifxundefined [1]{%
 \@ifx{#1\undefined}
}%
\providecommand \@ifnum [1]{%
 \ifnum #1\expandafter \@firstoftwo
 \else \expandafter \@secondoftwo
 \fi
}%
\providecommand \@ifx [1]{%
 \ifx #1\expandafter \@firstoftwo
 \else \expandafter \@secondoftwo
 \fi
}%
\providecommand \natexlab [1]{#1}%
\providecommand \enquote  [1]{``#1''}%
\providecommand \bibnamefont  [1]{#1}%
\providecommand \bibfnamefont [1]{#1}%
\providecommand \citenamefont [1]{#1}%
\providecommand \href@noop [0]{\@secondoftwo}%
\providecommand \href [0]{\begingroup \@sanitize@url \@href}%
\providecommand \@href[1]{\@@startlink{#1}\@@href}%
\providecommand \@@href[1]{\endgroup#1\@@endlink}%
\providecommand \@sanitize@url [0]{\catcode `\\12\catcode `\$12\catcode
  `\&12\catcode `\#12\catcode `\^12\catcode `\_12\catcode `\%12\relax}%
\providecommand \@@startlink[1]{}%
\providecommand \@@endlink[0]{}%
\providecommand \url  [0]{\begingroup\@sanitize@url \@url }%
\providecommand \@url [1]{\endgroup\@href {#1}{\urlprefix }}%
\providecommand \urlprefix  [0]{URL }%
\providecommand \Eprint [0]{\href }%
\providecommand \doibase [0]{https://doi.org/}%
\providecommand \selectlanguage [0]{\@gobble}%
\providecommand \bibinfo  [0]{\@secondoftwo}%
\providecommand \bibfield  [0]{\@secondoftwo}%
\providecommand \translation [1]{[#1]}%
\providecommand \BibitemOpen [0]{}%
\providecommand \bibitemStop [0]{}%
\providecommand \bibitemNoStop [0]{.\EOS\space}%
\providecommand \EOS [0]{\spacefactor3000\relax}%
\providecommand \BibitemShut  [1]{\csname bibitem#1\endcsname}%
\let\auto@bib@innerbib\@empty
\bibitem [{\citenamefont {Ringwald}(2014)}]{axionboson}%
  \BibitemOpen
  \bibfield  {author} {\bibinfo {author} {\bibfnamefont {A.}~\bibnamefont
  {Ringwald}},\ }\bibfield  {title} {\bibinfo {title} {Axions and axion-like
  particles},\ }\href@noop {} {\  (\bibinfo {year} {2014})},\ \Eprint
  {https://arxiv.org/abs/1407.0546} {arXiv:1407.0546 [hep-ph]} \BibitemShut
  {NoStop}%
\bibitem [{\citenamefont {Anselm}(1985)}]{anselm_arion-photon_1985}%
  \BibitemOpen
  \bibfield  {author} {\bibinfo {author} {\bibfnamefont {A.~A.}\ \bibnamefont
  {Anselm}},\ }\bibfield  {title} {\bibinfo {title} {{Arion $\leftrightarrow$
  Photon Oscillations in a Steady Magnetic Field. (In Russian)}},\ }\href@noop
  {} {\bibfield  {journal} {\bibinfo  {journal} {Yad. Fiz.}\ }\textbf {\bibinfo
  {volume} {42}},\ \bibinfo {pages} {1480} (\bibinfo {year}
  {1985})}\BibitemShut {NoStop}%
\bibitem [{\citenamefont {Van~Bibber}\ \emph {et~al.}(1987)\citenamefont
  {Van~Bibber}, \citenamefont {Dagdeviren}, \citenamefont {Koonin},
  \citenamefont {Kerman},\ and\ \citenamefont {Nelson}}]{Bibber_lsw}%
  \BibitemOpen
  \bibfield  {author} {\bibinfo {author} {\bibfnamefont {K.}~\bibnamefont
  {Van~Bibber}}, \bibinfo {author} {\bibfnamefont {N.~R.}\ \bibnamefont
  {Dagdeviren}}, \bibinfo {author} {\bibfnamefont {S.~E.}\ \bibnamefont
  {Koonin}}, \bibinfo {author} {\bibfnamefont {A.~K.}\ \bibnamefont {Kerman}},\
  and\ \bibinfo {author} {\bibfnamefont {H.~N.}\ \bibnamefont {Nelson}},\
  }\bibfield  {title} {\bibinfo {title} {Proposed experiment to produce and
  detect light pseudoscalars},\ }\href
  {https://doi.org/10.1103/PhysRevLett.59.759} {\bibfield  {journal} {\bibinfo
  {journal} {Phys. Rev. Lett.}\ }\textbf {\bibinfo {volume} {59}},\ \bibinfo
  {pages} {759} (\bibinfo {year} {1987})}\BibitemShut {NoStop}%
\bibitem [{\citenamefont {Bähre}\ \emph {et~al.}(2013)\citenamefont {Bähre},
  \citenamefont {Döbrich}, \citenamefont {Dreyling-Eschweiler}, \citenamefont
  {Ghazaryan}, \citenamefont {Hodajerdi}, \citenamefont {Horns}, \citenamefont
  {Januschek}, \citenamefont {Knabbe}, \citenamefont {Lindner}, \citenamefont
  {Notz}, \citenamefont {Ringwald}, \citenamefont {Seggern}, \citenamefont
  {Stromhagen}, \citenamefont {Trines},\ and\ \citenamefont
  {Willke}}]{alpsiitechnicaldesign}%
  \BibitemOpen
  \bibfield  {author} {\bibinfo {author} {\bibfnamefont {R.}~\bibnamefont
  {Bähre}}, \bibinfo {author} {\bibfnamefont {B.}~\bibnamefont {Döbrich}},
  \bibinfo {author} {\bibfnamefont {J.}~\bibnamefont {Dreyling-Eschweiler}},
  \bibinfo {author} {\bibfnamefont {S.}~\bibnamefont {Ghazaryan}}, \bibinfo
  {author} {\bibfnamefont {R.}~\bibnamefont {Hodajerdi}}, \bibinfo {author}
  {\bibfnamefont {D.}~\bibnamefont {Horns}}, \bibinfo {author} {\bibfnamefont
  {F.}~\bibnamefont {Januschek}}, \bibinfo {author} {\bibfnamefont {E.~A.}\
  \bibnamefont {Knabbe}}, \bibinfo {author} {\bibfnamefont {A.}~\bibnamefont
  {Lindner}}, \bibinfo {author} {\bibfnamefont {D.}~\bibnamefont {Notz}},
  \bibinfo {author} {\bibfnamefont {A.}~\bibnamefont {Ringwald}}, \bibinfo
  {author} {\bibfnamefont {J.~E.~v.}\ \bibnamefont {Seggern}}, \bibinfo
  {author} {\bibfnamefont {R.}~\bibnamefont {Stromhagen}}, \bibinfo {author}
  {\bibfnamefont {D.}~\bibnamefont {Trines}},\ and\ \bibinfo {author}
  {\bibfnamefont {B.}~\bibnamefont {Willke}},\ }\bibfield  {title} {\bibinfo
  {title} {Any light particle search {II} — technical design report},\ }\href
  {https://doi.org/10.1088/1748-0221/8/09/t09001} {\bibfield  {journal}
  {\bibinfo  {journal} {Journal of Instrumentation}\ }\textbf {\bibinfo
  {volume} {8}}\bibinfo  {number} { (09)},\ \bibinfo {pages}
  {T09001–T09001}}\BibitemShut {NoStop}%
\bibitem [{\citenamefont {Irastorza}\ and\ \citenamefont
  {Redondo}(2018)}]{alps2sensitivity}%
  \BibitemOpen
\bibfield  {number} {  }\bibfield  {author} {\bibinfo {author} {\bibfnamefont
  {I.~G.}\ \bibnamefont {Irastorza}}\ and\ \bibinfo {author} {\bibfnamefont
  {J.}~\bibnamefont {Redondo}},\ }\bibfield  {title} {\bibinfo {title} {New
  experimental approaches in the search for axion-like particles},\ }\href
  {https://doi.org/10.1016/j.ppnp.2018.05.003} {\bibfield  {journal} {\bibinfo
  {journal} {Progress in Particle and Nuclear Physics}\ }\textbf {\bibinfo
  {volume} {102}},\ \bibinfo {pages} {89–159} (\bibinfo {year}
  {2018})}\BibitemShut {NoStop}%
\bibitem [{\citenamefont {Ballou}\ \emph {et~al.}(2015)\citenamefont {Ballou}
  \emph {et~al.}}]{OSQAR:2015qdv}%
  \BibitemOpen
  \bibfield  {author} {\bibinfo {author} {\bibfnamefont {R.}~\bibnamefont
  {Ballou}} \emph {et~al.} (\bibinfo {collaboration} {OSQAR}),\ }\bibfield
  {title} {\bibinfo {title} {{New exclusion limits on scalar and pseudoscalar
  axionlike particles from light shining through a wall}},\ }\href
  {https://doi.org/10.1103/PhysRevD.92.092002} {\bibfield  {journal} {\bibinfo
  {journal} {Phys. Rev. D}\ }\textbf {\bibinfo {volume} {92}},\ \bibinfo
  {pages} {092002} (\bibinfo {year} {2015})},\ \Eprint
  {https://arxiv.org/abs/1506.08082} {arXiv:1506.08082 [hep-ex]} \BibitemShut
  {NoStop}%
\bibitem [{316720()}]{stellarevolution}%
  \BibitemOpen
  \bibfield  {author} {316720,\ }\bibfield  {title} {\bibinfo {title}
  {{P}roceedings, 11th {P}atras {W}orkshop on {A}xions, {WIMP}s and {WISP}s
  ({A}xion-{WIMP} 2015)},\ }\bibinfo {organization} {11th Patras Workshop on
  Axions, WIMPs and WISPs, Zaragoza (Spain), 22 Jun 2015 - 26 Jun 2015}\
  (\bibinfo  {publisher} {Verlag Deutsches Elektronen-Synchrotron},\ \bibinfo
  {address} {Hamburg},\ \bibinfo {year} {2015})\ pp.\ \bibinfo {pages}
  {1--251}\BibitemShut {NoStop}%
\bibitem [{\citenamefont {Ortiz}\ \emph {et~al.}(2022)\citenamefont {Ortiz}
  \emph {et~al.}}]{alps2counts}%
  \BibitemOpen
  \bibfield  {author} {\bibinfo {author} {\bibfnamefont {M.~D.}\ \bibnamefont
  {Ortiz}} \emph {et~al.},\ }\bibfield  {title} {\bibinfo {title} {{Design of
  the {ALPS}~{II} optical system}},\ }\href
  {https://doi.org/10.1016/j.dark.2022.100968} {\bibfield  {journal} {\bibinfo
  {journal} {Phys. Dark Univ.}\ }\textbf {\bibinfo {volume} {35}},\ \bibinfo
  {pages} {100968} (\bibinfo {year} {2022})},\ \Eprint
  {https://arxiv.org/abs/2009.14294} {arXiv:2009.14294 [physics.optics]}
  \BibitemShut {NoStop}%
\bibitem [{\citenamefont {Hallal}\ \emph {et~al.}(2022)\citenamefont {Hallal},
  \citenamefont {Messineo}, \citenamefont {Ortiz}, \citenamefont {Gleason},
  \citenamefont {Hollis}, \citenamefont {Tanner}, \citenamefont {Mueller},\
  and\ \citenamefont {Spector}}]{alps2het}%
  \BibitemOpen
  \bibfield  {author} {\bibinfo {author} {\bibfnamefont {A.}~\bibnamefont
  {Hallal}}, \bibinfo {author} {\bibfnamefont {G.}~\bibnamefont {Messineo}},
  \bibinfo {author} {\bibfnamefont {M.~D.}\ \bibnamefont {Ortiz}}, \bibinfo
  {author} {\bibfnamefont {J.}~\bibnamefont {Gleason}}, \bibinfo {author}
  {\bibfnamefont {H.}~\bibnamefont {Hollis}}, \bibinfo {author} {\bibfnamefont
  {D.}~\bibnamefont {Tanner}}, \bibinfo {author} {\bibfnamefont
  {G.}~\bibnamefont {Mueller}},\ and\ \bibinfo {author} {\bibfnamefont
  {A.}~\bibnamefont {Spector}},\ }\bibfield  {title} {\bibinfo {title} {The
  heterodyne sensing system for the alps ii search for sub-ev weakly
  interacting particles},\ }\href {https://doi.org/10.1016/j.dark.2021.100914}
  {\bibfield  {journal} {\bibinfo  {journal} {Physics of the Dark Universe}\
  }\textbf {\bibinfo {volume} {35}},\ \bibinfo {pages} {100914} (\bibinfo
  {year} {2022})}\BibitemShut {NoStop}%
\bibitem [{\citenamefont {Shah}\ \emph {et~al.}(2022)\citenamefont {Shah},
  \citenamefont {Isleif}, \citenamefont {Januschek}, \citenamefont {Lindner},\
  and\ \citenamefont {Schott}}]{epsrikhav}%
  \BibitemOpen
  \bibfield  {author} {\bibinfo {author} {\bibfnamefont {R.}~\bibnamefont
  {Shah}}, \bibinfo {author} {\bibfnamefont {K.-S.}\ \bibnamefont {Isleif}},
  \bibinfo {author} {\bibfnamefont {F.}~\bibnamefont {Januschek}}, \bibinfo
  {author} {\bibfnamefont {A.}~\bibnamefont {Lindner}},\ and\ \bibinfo {author}
  {\bibfnamefont {M.}~\bibnamefont {Schott}},\ }\bibfield  {title} {\bibinfo
  {title} {{TES} detector for {ALPS} {II}},\ }\bibfield  {journal} {\bibinfo
  {journal} {Proceedings of The European Physical Society Conference on High
  Energy Physics — PoS(EPS-HEP2021)}\ }\href
  {https://doi.org/10.22323/1.398.0801} {10.22323/1.398.0801} (\bibinfo {year}
  {2022})\BibitemShut {NoStop}%
\bibitem [{\citenamefont {Irwin}\ and\ \citenamefont
  {Hilton}(2005)}]{Irwin2005}%
  \BibitemOpen
  \bibfield  {author} {\bibinfo {author} {\bibfnamefont {K.}~\bibnamefont
  {Irwin}}\ and\ \bibinfo {author} {\bibfnamefont {G.}~\bibnamefont {Hilton}},\
  }\bibinfo {title} {Transition-edge sensors},\ in\ \href
  {https://doi.org/10.1007/10933596_3} {\emph {\bibinfo {booktitle} {Cryogenic
  Particle Detection}}}\ (\bibinfo  {publisher} {Springer Berlin Heidelberg},\
  \bibinfo {address} {Berlin, Heidelberg},\ \bibinfo {year} {2005})\ pp.\
  \bibinfo {pages} {63--150}\BibitemShut {NoStop}%
\bibitem [{\citenamefont {{Shah}}\ \emph {et~al.}(2022)\citenamefont {{Shah}},
  \citenamefont {{Isleif}}, \citenamefont {{Januschek}}, \citenamefont
  {{Lindner}},\ and\ \citenamefont {{Schott}}}]{jltprikhav}%
  \BibitemOpen
  \bibfield  {author} {\bibinfo {author} {\bibfnamefont {R.}~\bibnamefont
  {{Shah}}}, \bibinfo {author} {\bibfnamefont {K.-S.}\ \bibnamefont
  {{Isleif}}}, \bibinfo {author} {\bibfnamefont {F.}~\bibnamefont
  {{Januschek}}}, \bibinfo {author} {\bibfnamefont {A.}~\bibnamefont
  {{Lindner}}},\ and\ \bibinfo {author} {\bibfnamefont {M.}~\bibnamefont
  {{Schott}}},\ }\bibfield  {title} {\bibinfo {title} {{Characterising a
  Single-Photon Detector for {ALPS}~{II}}},\ }\href
  {https://doi.org/10.1007/s10909-022-02720-0} {\bibfield  {journal} {\bibinfo
  {journal} {Journal of Low Temperature Physics}\ }\textbf {\bibinfo {volume}
  {209}},\ \bibinfo {pages} {355} (\bibinfo {year} {2022})}\BibitemShut
  {NoStop}%
\bibitem [{\citenamefont {Rubiera~Gimeno}\ \emph
  {et~al.}(2023{\natexlab{a}})\citenamefont {Rubiera~Gimeno}, \citenamefont
  {Isleif}, \citenamefont {Januschek}, \citenamefont {Lindner}, \citenamefont
  {Meyer}, \citenamefont {Othman}, \citenamefont {Schott}, \citenamefont
  {Shah},\ and\ \citenamefont {Sohl}}]{rubiera_gimeno_tes_2023}%
  \BibitemOpen
  \bibfield  {author} {\bibinfo {author} {\bibfnamefont {J.~A.}\ \bibnamefont
  {Rubiera~Gimeno}}, \bibinfo {author} {\bibfnamefont {K.-S.}\ \bibnamefont
  {Isleif}}, \bibinfo {author} {\bibfnamefont {F.}~\bibnamefont {Januschek}},
  \bibinfo {author} {\bibfnamefont {A.}~\bibnamefont {Lindner}}, \bibinfo
  {author} {\bibfnamefont {M.}~\bibnamefont {Meyer}}, \bibinfo {author}
  {\bibfnamefont {G.}~\bibnamefont {Othman}}, \bibinfo {author} {\bibfnamefont
  {M.}~\bibnamefont {Schott}}, \bibinfo {author} {\bibfnamefont
  {R.}~\bibnamefont {Shah}},\ and\ \bibinfo {author} {\bibfnamefont
  {L.}~\bibnamefont {Sohl}},\ }\bibfield  {title} {\bibinfo {title} {The {TES}
  detector of the {ALPS} {II} experiment},\ }\href
  {https://doi.org/10.1016/j.nima.2022.167588} {\bibfield  {journal} {\bibinfo
  {journal} {Nuclear Instruments and Methods in Physics Research Section A:
  Accelerators, Spectrometers, Detectors and Associated Equipment}\ }\textbf
  {\bibinfo {volume} {1046}},\ \bibinfo {pages} {167588} (\bibinfo {year}
  {2023}{\natexlab{a}})}\BibitemShut {NoStop}%
\bibitem [{\citenamefont {Miller}\ \emph {et~al.}(2007)\citenamefont {Miller},
  \citenamefont {Lita}, \citenamefont {Rosenberg}, \citenamefont {Gruber},\
  and\ \citenamefont {Nam}}]{miller2007superconducting}%
  \BibitemOpen
  \bibfield  {author} {\bibinfo {author} {\bibfnamefont {A.~J.}\ \bibnamefont
  {Miller}}, \bibinfo {author} {\bibfnamefont {A.}~\bibnamefont {Lita}},
  \bibinfo {author} {\bibfnamefont {D.}~\bibnamefont {Rosenberg}}, \bibinfo
  {author} {\bibfnamefont {S.}~\bibnamefont {Gruber}},\ and\ \bibinfo {author}
  {\bibfnamefont {S.}~\bibnamefont {Nam}},\ }\bibfield  {title} {\bibinfo
  {title} {Superconducting photon number resolving detectors: {P}erformance and
  promise},\ }in\ \href@noop {} {\emph {\bibinfo {booktitle} {Proc. 8th Int.
  Conf. Quantum Communication, Measurement and Computing (QCMC’06)}}}\
  (\bibinfo {year} {2007})\ pp.\ \bibinfo {pages} {445--450}\BibitemShut
  {NoStop}%
\bibitem [{\citenamefont {Greiner}\ \emph {et~al.}(1995)\citenamefont
  {Greiner}, \citenamefont {Neise},\ and\ \citenamefont
  {Stöcker}}]{greiner_thermodynamics_1995}%
  \BibitemOpen
  \bibfield  {author} {\bibinfo {author} {\bibfnamefont {W.}~\bibnamefont
  {Greiner}}, \bibinfo {author} {\bibfnamefont {L.}~\bibnamefont {Neise}},\
  and\ \bibinfo {author} {\bibfnamefont {H.}~\bibnamefont {Stöcker}},\
  }\href@noop {} {\emph {\bibinfo {title} {Thermodynamics and statistical
  mechanics}}},\ Classical theoretical physics\ (\bibinfo  {publisher}
  {Springer},\ \bibinfo {year} {1995})\BibitemShut {NoStop}%
\bibitem [{\citenamefont {Lita}\ \emph {et~al.}(2010)\citenamefont {Lita},
  \citenamefont {Calkins}, \citenamefont {Pellouchoud}, \citenamefont
  {Miller},\ and\ \citenamefont {Nam}}]{lita_opt_stack}%
  \BibitemOpen
  \bibfield  {author} {\bibinfo {author} {\bibfnamefont {A.~E.}\ \bibnamefont
  {Lita}}, \bibinfo {author} {\bibfnamefont {B.}~\bibnamefont {Calkins}},
  \bibinfo {author} {\bibfnamefont {L.~A.}\ \bibnamefont {Pellouchoud}},
  \bibinfo {author} {\bibfnamefont {A.~J.}\ \bibnamefont {Miller}},\ and\
  \bibinfo {author} {\bibfnamefont {S.}~\bibnamefont {Nam}},\ }\bibfield
  {title} {\bibinfo {title} {{Superconducting transition-edge sensors optimized
  for high-efficiency photon-number resolving detectors}},\ }in\ \href
  {https://doi.org/10.1117/12.852221} {\emph {\bibinfo {booktitle} {Advanced
  Photon Counting Techniques IV}}},\ Vol.\ \bibinfo {volume} {7681},\ \bibinfo
  {editor} {edited by\ \bibinfo {editor} {\bibfnamefont {M.~A.}\ \bibnamefont
  {Itzler}}\ and\ \bibinfo {editor} {\bibfnamefont {J.~C.}\ \bibnamefont
  {Campbell}}},\ \bibinfo {organization} {International Society for Optics and
  Photonics}\ (\bibinfo  {publisher} {SPIE},\ \bibinfo {year} {2010})\ p.\
  \bibinfo {pages} {76810D}\BibitemShut {NoStop}%
\bibitem [{\citenamefont {Dreyling-Eschweiler}(2014)}]{janthesis}%
  \BibitemOpen
  \bibfield  {author} {\bibinfo {author} {\bibfnamefont {J.}~\bibnamefont
  {Dreyling-Eschweiler}},\ }\emph {\bibinfo {title} {{A} superconducting
  microcalorimeter for low-flux detection of near-infrared single photons}},\
  \href {https://doi.org/10.3204/DESY-THESIS-2014-016} {\bibinfo {type}
  {Dr.}},\ \bibinfo  {school} {University of Hamburg}, \bibinfo {address}
  {Hamburg} (\bibinfo {year} {2014}),\ \bibinfo {note} {university of Hamburg,
  Diss., 2014}\BibitemShut {NoStop}%
\bibitem [{\citenamefont {Arshad}\ \emph {et~al.}(2019)\citenamefont {Arshad},
  \citenamefont {Hartung},\ and\ \citenamefont {Jäger}}]{fiber_loss}%
  \BibitemOpen
  \bibfield  {author} {\bibinfo {author} {\bibfnamefont {M.~A.}\ \bibnamefont
  {Arshad}}, \bibinfo {author} {\bibfnamefont {A.}~\bibnamefont {Hartung}},\
  and\ \bibinfo {author} {\bibfnamefont {M.}~\bibnamefont {Jäger}},\
  }\bibfield  {title} {\bibinfo {title} {A stimulated stokes raman
  scattering-based approach for continuous wave supercontinuum generation in
  optical fibers},\ }\href {https://doi.org/10.1088/1612-202X/aaff53}
  {\bibfield  {journal} {\bibinfo  {journal} {Laser Physics Letters}\ }\textbf
  {\bibinfo {volume} {16}},\ \bibinfo {pages} {035108} (\bibinfo {year}
  {2019})}\BibitemShut {NoStop}%
\bibitem [{\citenamefont {Marcuse}(1982)}]{Simplified_Bending_loss_theory}%
  \BibitemOpen
  \bibfield  {author} {\bibinfo {author} {\bibfnamefont {D.}~\bibnamefont
  {Marcuse}},\ }\bibfield  {title} {\bibinfo {title} {Influence of curvature on
  the losses of doubly clad fibers},\ }\href
  {https://doi.org/10.1364/AO.21.004208} {\bibfield  {journal} {\bibinfo
  {journal} {Appl. Opt.}\ }\textbf {\bibinfo {volume} {21}},\ \bibinfo {pages}
  {4208} (\bibinfo {year} {1982})}\BibitemShut {NoStop}%
\bibitem [{\citenamefont {Smirnov}\ \emph {et~al.}(2015)\citenamefont
  {Smirnov}, \citenamefont {Vachtomin}, \citenamefont {Divochiy}, \citenamefont
  {Antipov},\ and\ \citenamefont {Goltsman}}]{Smirnov_2015}%
  \BibitemOpen
  \bibfield  {author} {\bibinfo {author} {\bibfnamefont {K.}~\bibnamefont
  {Smirnov}}, \bibinfo {author} {\bibfnamefont {Y.}~\bibnamefont {Vachtomin}},
  \bibinfo {author} {\bibfnamefont {A.}~\bibnamefont {Divochiy}}, \bibinfo
  {author} {\bibfnamefont {A.}~\bibnamefont {Antipov}},\ and\ \bibinfo {author}
  {\bibfnamefont {G.}~\bibnamefont {Goltsman}},\ }\bibfield  {title} {\bibinfo
  {title} {Dependence of dark count rates in superconducting single photon
  detectors on the filtering effect of standard single mode optical fibers},\
  }\href {https://doi.org/10.7567/APEX.8.022501} {\bibfield  {journal}
  {\bibinfo  {journal} {Applied Physics Express}\ }\textbf {\bibinfo {volume}
  {8}},\ \bibinfo {pages} {022501} (\bibinfo {year} {2015})}\BibitemShut
  {NoStop}%
\bibitem [{\citenamefont {Cabrera}\ \emph {et~al.}(1998)\citenamefont
  {Cabrera}, \citenamefont {Clarke}, \citenamefont {Colling}, \citenamefont
  {Miller}, \citenamefont {Nam},\ and\ \citenamefont
  {Romani}}]{cabrera_linear_tes}%
  \BibitemOpen
  \bibfield  {author} {\bibinfo {author} {\bibfnamefont {B.}~\bibnamefont
  {Cabrera}}, \bibinfo {author} {\bibfnamefont {R.~M.}\ \bibnamefont {Clarke}},
  \bibinfo {author} {\bibfnamefont {P.}~\bibnamefont {Colling}}, \bibinfo
  {author} {\bibfnamefont {A.~J.}\ \bibnamefont {Miller}}, \bibinfo {author}
  {\bibfnamefont {S.}~\bibnamefont {Nam}},\ and\ \bibinfo {author}
  {\bibfnamefont {R.~W.}\ \bibnamefont {Romani}},\ }\bibfield  {title}
  {\bibinfo {title} {{Detection of single infrared, optical, and ultraviolet
  photons using superconducting transition edge sensors}},\ }\href
  {https://doi.org/10.1063/1.121984} {\bibfield  {journal} {\bibinfo  {journal}
  {Applied Physics Letters}\ }\textbf {\bibinfo {volume} {73}},\ \bibinfo
  {pages} {735} (\bibinfo {year} {1998})}\BibitemShut {NoStop}%
\bibitem [{\citenamefont {Lita}\ \emph {et~al.}(2008)\citenamefont {Lita},
  \citenamefont {Miller},\ and\ \citenamefont {Nam}}]{enrgres}%
  \BibitemOpen
  \bibfield  {author} {\bibinfo {author} {\bibfnamefont {A.}~\bibnamefont
  {Lita}}, \bibinfo {author} {\bibfnamefont {A.}~\bibnamefont {Miller}},\ and\
  \bibinfo {author} {\bibfnamefont {S.}~\bibnamefont {Nam}},\ }\bibfield
  {title} {\bibinfo {title} {Energy collection efficiency of tungsten
  transition-edge sensors in the near-infrared},\ }\href
  {https://doi.org/10.1007/s10909-007-9627-z} {\bibfield  {journal} {\bibinfo
  {journal} {Journal of Low Temperature Physics}\ }\textbf {\bibinfo {volume}
  {151}},\ \bibinfo {pages} {125} (\bibinfo {year} {2008})}\BibitemShut
  {NoStop}%
\bibitem [{\citenamefont {Li}\ \emph {et~al.}(2023)\citenamefont {Li},
  \citenamefont {Zhong}, \citenamefont {Zhang}, \citenamefont {Wang},
  \citenamefont {Zhou}, \citenamefont {Miao}, \citenamefont {Ren},
  \citenamefont {Li}, \citenamefont {Yao},\ and\ \citenamefont
  {Shi}}]{multicolor}%
  \BibitemOpen
  \bibfield  {author} {\bibinfo {author} {\bibfnamefont {P.}~\bibnamefont
  {Li}}, \bibinfo {author} {\bibfnamefont {J.}~\bibnamefont {Zhong}}, \bibinfo
  {author} {\bibfnamefont {W.}~\bibnamefont {Zhang}}, \bibinfo {author}
  {\bibfnamefont {Z.}~\bibnamefont {Wang}}, \bibinfo {author} {\bibfnamefont
  {K.}~\bibnamefont {Zhou}}, \bibinfo {author} {\bibfnamefont {W.}~\bibnamefont
  {Miao}}, \bibinfo {author} {\bibfnamefont {Y.}~\bibnamefont {Ren}}, \bibinfo
  {author} {\bibfnamefont {J.}~\bibnamefont {Li}}, \bibinfo {author}
  {\bibfnamefont {Q.}~\bibnamefont {Yao}},\ and\ \bibinfo {author}
  {\bibfnamefont {S.}~\bibnamefont {Shi}},\ }\bibfield  {title} {\bibinfo
  {title} {Multi-color photon detection with a single superconducting
  transition-edge sensor},\ }\href
  {https://doi.org/https://doi.org/10.1016/j.nima.2023.168408} {\bibfield
  {journal} {\bibinfo  {journal} {Nuclear Instruments and Methods in Physics
  Research Section A: Accelerators, Spectrometers, Detectors and Associated
  Equipment}\ }\textbf {\bibinfo {volume} {1054}},\ \bibinfo {pages} {168408}
  (\bibinfo {year} {2023})}\BibitemShut {NoStop}%
\bibitem [{\citenamefont {Eckart}\ and\ \citenamefont
  {Shonka}(1938)}]{Pileup_BBR}%
  \BibitemOpen
  \bibfield  {author} {\bibinfo {author} {\bibfnamefont {C.}~\bibnamefont
  {Eckart}}\ and\ \bibinfo {author} {\bibfnamefont {F.~R.}\ \bibnamefont
  {Shonka}},\ }\bibfield  {title} {\bibinfo {title} {Accidental coincidences in
  counter circuits},\ }\href {https://doi.org/10.1103/PhysRev.53.752}
  {\bibfield  {journal} {\bibinfo  {journal} {Phys. Rev.}\ }\textbf {\bibinfo
  {volume} {53}},\ \bibinfo {pages} {752} (\bibinfo {year} {1938})}\BibitemShut
  {NoStop}%
\bibitem [{\citenamefont {Rubiera~Gimeno}\ \emph
  {et~al.}(2023{\natexlab{b}})\citenamefont {Rubiera~Gimeno}, \citenamefont
  {Januschek}, \citenamefont {Isleif}, \citenamefont {Lindner}, \citenamefont
  {Meyer}, \citenamefont {Othman}, \citenamefont {Schwemmbauer},\ and\
  \citenamefont {Shah}}]{RubieraGimeno_eps}%
  \BibitemOpen
  \bibfield  {author} {\bibinfo {author} {\bibfnamefont {J.~A.}\ \bibnamefont
  {Rubiera~Gimeno}}, \bibinfo {author} {\bibfnamefont {F.}~\bibnamefont
  {Januschek}}, \bibinfo {author} {\bibfnamefont {K.-S.}\ \bibnamefont
  {Isleif}}, \bibinfo {author} {\bibfnamefont {A.}~\bibnamefont {Lindner}},
  \bibinfo {author} {\bibfnamefont {M.}~\bibnamefont {Meyer}}, \bibinfo
  {author} {\bibfnamefont {G.}~\bibnamefont {Othman}}, \bibinfo {author}
  {\bibfnamefont {C.}~\bibnamefont {Schwemmbauer}},\ and\ \bibinfo {author}
  {\bibfnamefont {R.}~\bibnamefont {Shah}},\ }\bibfield  {title} {\bibinfo
  {title} {{A TES system for ALPS II - Status and Prospects}},\ }\href
  {https://doi.org/10.22323/1.449.0567} {\bibfield  {journal} {\bibinfo
  {journal} {PoS}\ }\textbf {\bibinfo {volume} {EPS-HEP2023}},\ \bibinfo
  {pages} {567} (\bibinfo {year} {2023}{\natexlab{b}})}\BibitemShut {NoStop}%
\bibitem [{\citenamefont {Meyer}\ \emph {et~al.}(2023)\citenamefont {Meyer},
  \citenamefont {Othman}, \citenamefont {Isleif}, \citenamefont {Januschek},
  \citenamefont {Lindner}, \citenamefont {Rubiera~Gimeno}, \citenamefont
  {Schwemmbauer}, \citenamefont {Schott},\ and\ \citenamefont
  {Shah}}]{manuel_ML}%
  \BibitemOpen
  \bibfield  {author} {\bibinfo {author} {\bibfnamefont {M.}~\bibnamefont
  {Meyer}}, \bibinfo {author} {\bibfnamefont {G.}~\bibnamefont {Othman}},
  \bibinfo {author} {\bibfnamefont {K.}~\bibnamefont {Isleif}}, \bibinfo
  {author} {\bibfnamefont {F.}~\bibnamefont {Januschek}}, \bibinfo {author}
  {\bibfnamefont {A.}~\bibnamefont {Lindner}}, \bibinfo {author} {\bibfnamefont
  {J.~A.}\ \bibnamefont {Rubiera~Gimeno}}, \bibinfo {author} {\bibfnamefont
  {C.}~\bibnamefont {Schwemmbauer}}, \bibinfo {author} {\bibfnamefont
  {M.}~\bibnamefont {Schott}},\ and\ \bibinfo {author} {\bibfnamefont
  {R.}~\bibnamefont {Shah}},\ }\bibfield  {title} {\bibinfo {title} {{A first
  application of machine and deep learning for background rejection in the ALPS
  II TES detector}},\ }\bibfield  {booktitle} {\emph {\bibinfo {booktitle}
  {{17th Workshop on Axions, WIMPs and WISPs}}}\ }\href
  {https://doi.org/10.1002/andp.202200545} {10.1002/andp.202200545} (\bibinfo
  {year} {2023}),\ \Eprint {https://arxiv.org/abs/2304.08406} {arXiv:2304.08406
  [hep-ex]} \BibitemShut {NoStop}%
\end{thebibliography}%

\end{document}